\documentclass[a4paper, 12pt]{article}
\usepackage{epsfig,amssymb,euscript,xspace, color}
\usepackage{amsmath,jheppub,mathtools,empheq,amsthm}
\usepackage{graphicx}
\usepackage{verbatim}
\usepackage{cancel}
\usepackage[T1]{fontenc} 
\usepackage{tikz,caption,subcaption,marvosym} 
\usepackage{amsfonts}
\usetikzlibrary{decorations.markings,arrows,snakes}

\def\bea{\begin{eqnarray}}
\def\eea{\end{eqnarray}}
\def\be{\begin{equation}}
\def\ee{\end{equation}}

\newcommand{\tr}{\text{Tr }}

\definecolor{lightblue}{rgb}{.1,.4,.5}
\definecolor{brown1}{rgb}{.64,.43,.138}



\title{ $\widehat{sl_2}$ symmetry of ${\mathbb R}^{1,3}$ gravity}
\author{Nishant Gupta$^{a,b}$, Partha Paul$^c$ and Nemani V. Suryanarayana$^{a,b}$}
\emailAdd{nishantg@imsc.res.in, nemani@imsc.res.in, parthapaul@cmi.ac.in}
\affiliation{$^a$Institute of Mathematical Sciences, \\ Taramani, Chennai 600 113, India \\ \\ $^b$Homi Bhabha National Institute,  \\ Anushakti Nagar, Mumbai 400094, India \\ \\
$^c$Chennai Mathematical Institute, \\ SIPCOT IT Park, Siruseri, Chennai 603103, India}

\abstract{We propose novel asymptotically locally flat boundary conditions for Einstein Gravity without cosmological constant in four dimensions that are consistent with the variational principle. They allow for complex solutions that are asymptotically diffeomorphic to flat space-times under complexified diffeomorphisms. We show that the resultant asymptotic symmetries are an extension of the Poincare algebra to a copy of Virasoro, a chiral $\mathfrak{sl}(2,{\mathbb C})$ current algebra along with two chiral $\mathfrak{u}(1)$ currents. We posit that these bulk symmetries are direct analogues of the recently discovered chiral algebra symmetries of gravitational scattering amplitudes as celestial CFT correlation functions. }

\begin{document}
\maketitle

\section{Introduction}

The holographic description of any 4-dimensional gravitational theory  with asymptotically locally flat  boundary conditions is one of the most interesting questions being pursued in recent times. At the future or the past null infinity the boundary is a three dimensional manifold with a degenerate metric and the holographic theory is expected to include sectors residing at these null infinities. With Dirichlet type boundary conditions where the non-degenerate part of the boundary metric is taken to be that of a round $S^2$  the asymptotic symmetry algebra is the famous $\mathfrak{bms}_4$ \cite{Bondi:1962bm, Sachs:1962sa, Sachs:1962rs} algebra that includes the Poincare algebra as its maximal finite-dimensional sub-algebra. In fact allowing for singular vector fields on $S^2$ extends the $\mathfrak{bms}_4$  to the so-called extended $\mathfrak{bms}_4$ algebra where the Lorentz transformation algebra $\mathfrak{sl}_2 \oplus \overline {\mathfrak{sl}_2}$ becomes two copies of Virasoro algebra \cite{barnich:2010tr, barnich:2010tro, barnich:2010ct, barnich:2011hc}. A different set of boundary conditions considered in \cite{campiglia:2014la, campiglia:2015al} produced a different extension called generalised $\mathfrak{bms}_4$ that extends  $\mathfrak{sl}_2 \oplus \overline {\mathfrak{sl}_2}$ to all smooth diffeomorphisms of $S^2$. These two extensions have been shown \cite{kapec:2014lps, campiglia:2014la} to be responsible for the sub-leading soft graviton theorems \cite{Cachazo:2014fwa}.  

However, it is fair to say it is unclear what the maximal set of symmetries admitted by ${\mathbb R}^{1,3}$ gravity is. There is another and more urgent motivation to consider this question. Recently 
in \cite{banerjee:2020gp, Banerjee:2021dlm} a novel $\mathfrak{sl}_2$ current algebra symmetry has been uncovered as the symmetry consistent with the OPEs of graviton operators in the celestial CFT. How such a symmetry algebra can be seen directly from classical gravity is an important question that we address in this paper. In particular, we seek boundary conditions for the ${\mathbb R}^{1,3}$ gravity whose asymptotic symmetry algebra includes the $\mathfrak{sl}_2$ current algebra of \cite{banerjee:2020gp, Banerjee:2021dlm}. 

 Our analysis is for Locally Flat solutions in $\mathbb{R}^{1,3}$ gravity, where motivated by chiral gravity boundary conditions for $AdS_3$-gravity considered in \cite{Avery:2013dja, Apolo:2014tua}, we allow for chiral fluctuations of the metric on the spatial part ($\Sigma_2$) of the null infinity (celestial sphere or celestial plane). These are non-radiative solutions on which we further impose variational principle with appropriate boundary action akin to $AdS_3$-gravity case. The corresponding asymptotic symmetry analysis of such locally flat solutions leads to a novel extension of the Poincare algebra to an infinite dimensional algebra that consists of one copy of the Virasoro, one copy of the $\mathfrak{sl}_2$ current algebra and two $\mathfrak{u}(1)$ current algebras which match with the symmetry algebra uncovered from the celestial CFT in \cite{banerjee:2020gp, Banerjee:2021dlm}. The resultant non-linear solutions we obtain after imposing the variational principle are characterised by Goldstone modes associated with the spontaneous breaking of this symmetry algebra in the gravitational vacuum.
The results of our analysis include,
\begin{itemize}
\item Providing all locally flat solutions (including complex ones) with time-independent boundary metrics on $\Sigma_2$.
\item Using a set of appropriate boundary terms extracting further conditions on the space of locally flat solutions coming from the variational principle.
\item Choosing a chiral gauge for the metric on $\Sigma_2$ showing that the residual large diffeomorphisms generate the desired infinite dimensional chiral algebra that includes an $\mathfrak{sl}_2$ current algebra.
\end{itemize}
The rest of the paper is organised as follows. In Section \ref{sec2} we review the relevant aspects of the two directions that have led us to our investigation, namely, the hidden symmetries 2$d$ gravity of Polyakov from the $AdS_3$ perspective, and the recently uncovered chiral current algebra symmetries of the celestial CFT. In Section \ref{sec3} we consider construction of asymptotically locally flat and exactly locally flat solutions of interest. In Section \ref{sec4} the variational problem arising from the addition of the Gibbons-Hawking like boundary terms to ${\mathbb R}^{1,3}$ gravity with a cut-off near future (past) null infinity is analysed. In Section \ref{sec5} we carry out the asymptotic symmetry analysis of the solution spaces that solve the variational problem and compute the desired algebras. We provide a discussion of our results and open questions in Section \ref{sec6}.

\section{A motivation and a review}
\label{sec2}
In this section we review two aspects that act as motivation and guidance  for the rest of our paper: The emergence of  $(i)$ $\mathfrak{sl}_2$ current algebra as part of the symmetries of the $2d$ celestial CFT, $(ii)$
$\mathfrak{sl}_2$ current algebra in $AdS_3$ gravity with chiral gravity boundary conditions. 
\subsection{Current Algebra from Celestial CFT}
The $S$-matrix elements of ${\mathbb R}^{1,3}$ gravity can be recast as $2d$ conformal correlators, called the celestial amplitudes \cite{pasterski:2017ss,pasterski:2017sh}. By taking conformal soft limits one can uncover Ward identities of various $2d$ conformal currents -- referred to as the conformal soft theorems. Here we review briefly the existence of two current algebra symmetries of the $2d$ celestial amplitudes that follow from the leading and subleading conformal soft theorems for positive helicity soft graviton operators (see \cite{banerjee:2020gp} for notation and more details). For this one starts with the leading order conformal soft theorem for an outgoing positive helicity soft graviton operator $ S^+_0(z,\bar{z}) $
\be\label{ws}
\big\langle{S^+_0(z,\bar{z})} \prod_{i=1}^n \phi_{h_i,\bar{h}_i}(z_i,\bar{z}_i)\big\rangle = - \left(\sum_{k=1}^n \frac{\bar{z} - \bar{z}_k}{z - z_k} \epsilon_k \mathcal{P}_k\right) \big\langle{\prod_{i=1}^n \phi_{h_i,\bar{h}_i}(z_i,\bar{z}_i)}\big\rangle
\ee
where  $  \mathcal{P}_k \phi_{h_i,\bar{h}_i}(z_i,\bar{z}_i) = \delta_{ki}  \phi_{h_i+1/2,\bar{h}_i+1/2}(z_i,\bar{z}_i)$, $ \epsilon_k = \pm 1 $ for an outgoing (incoming) particle. Also $(h_k,\bar h_k) = (\frac{\Delta_k+\sigma_k}{2}, \frac{\Delta_k -\sigma_k}{2} )$ where $ \Delta_k$ and $\sigma_k $ are the scaling dimension and helicity of the $ k $-th particle respectively. The RHS of equation \eqref{ws} is a polynomial in $ \bar z $, so we can expand it around $ \bar{z}=0 $ and rewrite the equation as
\be\label{wis}
\begin{gathered}
\big\langle{S^+_0(z,\bar z)} \prod_{i=1}^n \phi_{h_i,\bar h_i}(z_i,\bar z_i)\big\rangle \\
= \left(\sum_{k=1}^n \frac{\bar z_k}{z - z_k} \epsilon_k \mathcal{P}_k\right) \big\langle{\prod_{i=1}^n \phi_{h_i,\bar h_i}(z_i,\bar z_i)}\big\rangle - \bar z \left(\sum_{k=1}^n \frac{1}{z - z_k} \epsilon_k \mathcal{P}_k\right) \big\langle{\prod_{i=1}^n \phi_{h_i,\bar h_i}(z_i,\bar z_i)}\big\rangle
\end{gathered}
\ee
Now if we define two currents $C^{\frac{1}{2}}(z)$ and $C^{-\frac{1}{2}}(z)$ in the following way,
\be
S^+_0(z,\bar z) = C^{\frac{1}{2}}(z) - \bar z \, C^{-\frac{1}{2}}(z)
\ee
then equation \eqref{wis} implies separate Ward identities of these two currents given by, 
\be\label{c1}
\big\langle{C^{\frac{1}{2}}(z) \prod_{i=1}^n \phi_{h_i,\bar h_i}(z_i,\bar z_i)}\big\rangle = \left(\sum_{k=1}^n \frac{\bar z_k}{z - z_k} \epsilon_k \mathcal{P}_k\right) \big\langle{\prod_{i=1}^n \phi_{h_i,\bar h_i}(z_i,\bar z_i)}\big\rangle
\ee
and 
\be\label{c2}
\big\langle{C^{-\frac{1}{2}}(z) \prod_{i=1}^n \phi_{h_i,\bar h_i}(z_i,\bar z_i)}\big\rangle = \left(\sum_{k=1}^n \frac{1}{z - z_k}\epsilon_k \mathcal{P}_k\right) \big\langle{\prod_{i=1}^n \phi_{h_i,\bar h_i}(z_i,\bar z_i)}\big\rangle
\ee
One can see from \eqref{c1} and \eqref{c2} that $C^{\frac{1}{2}}(z)$ and $C^{-\frac{1}{2}}(z)$ are generators of infinitesimal supertranslations acting as:
\be
\delta_{P^{\frac{1}{2}}}\phi_{h,\bar h}(z,\bar z) = \, \bar z  \,P^{\frac{1}{2}} (z) \, \epsilon \mathcal{P} \, \phi_{h,\bar h}(z,\bar z)
\ee
and
\be
\delta_{P^{-\frac{1}{2}}}\phi_{h,\bar h}(z,\bar z) = P^{-\frac{1}{2}}(z) \, \epsilon \mathcal{P} \phi_{h,\bar h}(z,\bar z)
\ee
respectively. 
Next we turn to the emergence of $ \mathfrak{sl}_2 $ current algebra from the celestial CFT. For this one starts with the corresponding subleading conformal soft graviton theorem (the holographic analog of \cite{Cachazo:2014fwa}) for a positive helicity outgoing soft graviton operator $S^+_1(z,\bar z) $
\be\label{ss}
\big\langle{S^+_1(z,\bar z)} \prod_{i=1}^n \phi_{h_i,\bar h_i}(z_i,\bar z_i)\big\rangle =  \sum_{k=1}^n \frac{(\bar z- \bar z_k)^2}{z-z_k}  \left[\frac{2\bar h_k}{\bar z- \bar z_k} - \bar\partial_k\right] \langle{\prod_{i=1}^n \phi_{h_i, \bar h_i}(z_i,\bar z_i)}\rangle
\ee
where $S^+_1(z,\bar z) $ is given by
\be
S^+_1(z,\bar z) = \lim_{\Delta\rightarrow 0} \Delta \, G^+_{\Delta}(z,\bar z)
\ee
One follows the same procedure as leading conformal soft theorem above. Expanding the RHS of \eqref{ss} in powers of $\bar z$ gives
\be
\begin{gathered}
\big\langle{S^+_1(z,\bar z)} \prod_{i=1}^n \phi_{h_i,\bar h_i}(z_i,\bar z_i)\big\rangle \\ = - \sum_{k=1}^n \frac{\bar z_k^2 \bar\partial_k + 2\bar h_k \bar z_k}{z- z_k} \ \langle{\prod_{i=1}^n \phi_{h_i,\bar h_i}(z_i,\bar z_i)}\rangle + 2 \bar z \sum_{k=1}^n \frac{\bar h_k + \bar z_k \bar\partial_k}{z- z_k} \ \langle{\prod_{i=1}^n \phi_{h_i,\bar h_i}(z_i,\bar z_i)}\rangle \\ - \bar z^2 \sum_{k=1}^n \frac{1}{z- z_k} \frac{\partial}{\partial \bar z_k} \ \langle{\prod_{i=1}^n \phi_{h_i,\bar h_i}(z_i,\bar z_i)}\rangle 
\end{gathered}
\ee
Using this one can define three currents $J^i(z)$ where $i= 0, \pm 1$, which are the generators of ${\mathfrak{sl}_2}$. In terms of these currents we can write the soft graviton operator $S^+_1(z,\bar z)$ as, 
\be
S^+_1(z,\bar z) = - J^{1}(z) + 2 \, \bar z \, J^{0}(z) - \bar z^2 J^{-1}(z) 
\ee 
The mode algebra of the currents $J^i(z)$ is
\be\label{g0}
\left[J^i_m, J^j_n\right] = (i-j) J^{i+j}_{m+n}
\ee
The symmetry algebra we have discussed so far has been obtained by analysing the correlation functions between the primary operators of the $2d$ celestial field theory and conformally soft gravitons. If a holographic duality exists between gravitational theories in four dimensional asymptotically flat spacetime and $2d$ celestial conformal field theory then the bulk theory should also posses these symmetries. This motivates us to search for these symmetries directly in the bulk ${\mathbb R}^{1,3}$ gravity. 

Next we turn to an older computation \cite{Avery:2013dja} in the context of $AdS_3$ gravity that led to a similar $\mathfrak{sl}_2$ current algebra. We will use an almost identical computation in ${\mathbb R}^{1,3}$ gravity to uncover the desired symmetry algebra.

\subsection{An $\widehat{\mathfrak{sl}_2}$ algebra from $AdS_3$ gravity}
An $\mathfrak{sl}_2$ current algebra can be seen as the asymptotic symmetry algebra $AdS_3$ gravity by taking the boundary metric of asymptotically $AdS_3$ geometries in the Polyakov gauge \cite{Polyakov:1987zb}. Working in the Fefferman-Graham gauge the locally $AdS_3$ geometries can be written as
\bea
ds_{lAdS_3}^2 =  l^2 \, \frac{dr^2}{r^2} +  r^2 \, \left( g^{(0)}_{ab} + \frac{l^2}{r^2}  g^{(2)}_{ab} + \frac{l^4}{r^4} g^{(4)}_{ab} \right) dx^a dx^b
\eea
along with 
\bea
\label{ads3-conds}
g^{(4)}_{ab} = \frac{1}{4} g^{(2)}_{ac} g_{(0)}^{cd} g^{(2)}_{db}, ~~~ \nabla_{(0)}^a g^{(2)}_{ab} = \frac{1}{2} \partial_b R_0, ~~ {\rm and} ~~ g^{ab}_{(0)} g^{(2)}_{ab} = \frac{1}{2} R_{(0)}.
\eea
%
Choosing the chiral gravity gauge of Polyakov 
\bea
\label{polyakovgauge}
ds^2=  (F(z,\bar z) \, dz + d\bar z) \, dz
\eea
for the boundary metric $g^{(0)}_{ab}$ the differential conditions impose the following equation:
\bea
\left(\partial_z g^{(2)}_{\bar z \bar z} - F \, \partial_{\bar z} - 2 \, \partial_{\bar z} F \right) \, g^{(2)}_{\bar z\bar z} = \frac{1}{2} \partial_{\bar z}^3 F \, .
\eea
One also has
\bea
g^{(2)}_{zz} = \kappa(z) + F^2 \, g^{(2)}_{\bar z \bar z} + \frac{1}{4} \left[ 
(\partial_{\bar z} F)^2 + 2 \, F \, \partial_{\bar z}^2 F +2 \, \partial_z \partial_{\bar z} F \right] \eea
%
%
Finally the bulk variational problem can be satisfied by setting $g^{(2)}_{\bar z \bar z} =0$ as shown in \cite{Avery:2013dja}. This makes $F(z, \bar z)$ a polynomial of degree 2 in $\bar z$. The residual bulk diffeomorphisms lead to one copy of $\mathfrak{sl}(2, {\mathbb R})$ current algebra and one copy of Virasoro algebra.\footnote{Strictly speaking the $\mathfrak{sl}(2, {\mathbb R})$ current algebra is the correct one only when one is talking about the Lorentizan $AdS_3$ theory. When dealing with Euclidean theory we should say the relevant algebra is $\mathfrak{sl}(2, {\mathbb C})$.}

In what follows, we implement a similar procedure in ${\mathbb R}^{1,3}$ gravity and demonstrate that an analogous $\mathfrak{sl}_2$ chiral current algebra emerges in this context as well.

\section{A class of ALF spacetimes in four dimensions}
\label{sec3}
We start with collecting some useful formulae towards construction of asymptotically locally flat (ALF) solutions in ${\mathbb R}^{1,3}$ gravity. We find it convenient to use the Newman-Unti gauge \cite{Barnich:2011ty} and work with the coordinates $(r, u, z,\bar z)$ suitable for the future null infinity. Then we have
\bea
g_{rr} =g_{rz} = g_{r\bar z}=0, ~~~ g_{ur} =-1 .
\eea
Let us write the remaining metric components $g_{ij}$ for $i,j \in \{ u, z, \bar z\}$ in the following form:
\bea
g_{ij} (r, u, z, \bar z) &=& \sum_{n=0}^\infty r^{2-n} \, g_{ij}^{(n)}(u,z, \bar z) 
\eea
We seek Ricci flat metrics ($R_{\mu\nu} =0$) which are also asymptotically locally flat. The latter condition is implemented by demanding that ${R^\mu}_{\nu\sigma\lambda} \rightarrow 0$ as $r \rightarrow \infty$ \cite{Ashtekar:2014zsa}. 

We first solve for $R_{\mu\nu} =0$ for large $r$ in a power series in $1/r$. The leading non-trivial conditions require that $g^{(0)}_{ij}$ is degenerate. Since we are interested in solutions for which the metric $g^{(0)}_{ab}$ for $a,b \in \{z, \bar z\}$ on the spatial manifold $\Sigma_2$ is non-degenerate we solve this condition by assuming:
\bea
\label{confconds1}
g^{(0)}_{uu} = g^{(0)}_{uz} =g^{(0)}_{u\bar z} =0 \, .
\eea
Then the next non-trivial condition implies that $g_{uu}^{(1)} = - \frac{1}{2} \partial_u ({\rm log} \det g^{(0)}_{ab})$. One also finds the condition $\det \, (\hat \gamma) = \frac{1}{4} ({\rm tr} \, \hat \gamma)^2$, for the $2 \times 2$ matrix $\hat \gamma^a_b = g_{(0)}^{ac} \partial_u g^{(0)}_{cb}$. Any $2 \times 2$ matrix that satisfies such relation has to have both its eigenvalues the same -- and therefore can only be proportional to the $2\times 2$ identity matrix $I_{2}$. Thus we arrive at the condition:
$\partial_u g^{(0)}_{ab} = \lambda \, g^{(0)}_{ab}$  $\forall a,b \in \{z, \bar z\}$ with the same $\lambda$. This immediately leads to $g^{(0)}_{ab} (u, z, \bar z) = \Omega (u, z, \bar z) q_{ab} (z, \bar z)$ where $q_{ab}$ is $u$ independent general $2\times 2 $ matrix. At ${\cal O}(1/r)$ from vanishing of $R_{uz}$ and $R_{u\bar z}$ we find 
%
$g^{(1)}_{uz} = g^{(1)}_{u\bar z} =0$.
%
We will further assume that the conformal factor $\Omega$ for the boundary metric $g^{(0)}_{ab}$ is independent of $u$ which in turn implies $g_{uu}^{(1)}=0$. Further one finds:
\bea
g^{(2)}_{uu} = - \frac{1}{2} R - \frac{1}{2} \partial_u \left(g_{(0)}^{ab} g^{(1)}_{ab} \right)   \, ,
\eea
\bea
g^{(2)}_{ua} = V_{a}(z,\bar z) + \frac{1}{2} D^b g^{(1)}_{ba} - \frac{1}{2 } D_a \left(g_{(0)}^{cd} 
g^{(1)}_{cd} \right) 
\eea
and
\bea
g^{(2)}_{ab} (u, z, \bar z) = D_{ab}( z, \bar z) + \frac{1}{4} g^{(1)}_{ac} \, g_{(0)}^{cd} \, g^{(1)}_{cb}  \, .
\eea
and so on, where $V_a(z,\bar{z})$ and $D_{ab}(z,\bar{z})$ are unconstrained $u$-independent $2d$ vector and rank--2 symmetric tensor respectively.  It turns out that imposing vanishing of ${R^\mu}_{\nu\sigma\lambda}$ at order $r$ implies
\bea
\label{confconds3}
\partial_u^2 \left[ g^{(1)}_{ab} - \frac{1}{2} \left(g_{(0)}^{cd} g^{(1)}_{cd} \right) \, g^{(0)}_{ab} \right] = 0 \, .
\eea
%
%
%
%
This means that one can determine the $u$-dependence of the trace-free part of $g^{(1)}_{ab}$ completely and it is at most linear in $u$. Any such solution can be written as 
\bea
g^{(1)}_{ab} - \frac{1}{2} \left(g_{(0)}^{cd} g^{(1)}_{cd} \right) \, g^{(0)}_{ab} &=& g^{(1,0)}_{ab} (z, \bar z) + (u-u_0) \, g^{(1,1)}_{ab} (z, \bar z)
\eea
where \{$g^{(1,0)}_{ab} (z, \bar z), g^{(1,1)}_{ab} (z, \bar z)$\} are traceless and symmetric.  
The remaining data in $(g^{(0)}_{ab}, g^{(1)}_{ab} )$ has to satisfy some further differential conditions. In particular:
\bea
\label{Tab1}
D^b \left(\partial_u \left[g^{(1)}_{ba} - \frac{1}{2}\left(g_{(0)}^{cd} g^{(1)}_{cd} \right) \, g^{(0)}_{ba} \right] + \frac{1}{2} R_0 \, g^{(0)}_{ba} \right)=0 
\eea
%
%
where $R_{(0)}$ is the scalar curvature of the 2$d$ metric $g^{(0)}_{ab}$. 
%
This equation (\ref{Tab1}) is similar to that in (\ref{ads3-conds})   of $AdS_3$ gravity. 
%

Further constraints arising from asymptotic flatness are: $V_{a}( z, \bar z)=0$, $\partial_u g_{uu}^{(3)} (u, z, \bar z) =0$. Along with $\partial_a g_{uu}^{(3)}$ determined in terms of the data at order $r^2$ and order $r$ and differential conditions on $D_{ab}$. More unconstrained data also appears at ${\cal O}(r^{-1})$ and Ricci flat solutions can be systematically constructed order by order in powers of $r^{-1}$. It turns out that $g_{(0)}^{ab}g^{(1)}_{ab}$ remains unconstrained and we set it to zero as a further gauge condition \cite{Barnich:2011ty}.

However, to study the asymptotic symmetries of these solutions one does not need to find these solution spaces completely. It is sufficient to find the  locally flat (LF) solutions that share the boundary conditions derived so far.
\subsection{Classes of locally flat geometries}
These form a subset of Ricci flat geometries discussed above which have vanishing Riemann tensors. With our NU gauge and boundary conditions we find that any such locally flat solution can be written as polynomials in $r$:
\bea
\label{lf-configs}
ds^2_{LF} =  g^{(2)}_{uu} \, du^2 -2 \, du \, dr + \left(r^2 \, g^{(0)}_{ab} + r g^{(1)}_{ab} + g^{(2)}_{ab} \right) \, dx^a \, dx^b + 2 \, g^{(2)}_{ua} \, dx^a \, du
\eea
where 
\bea
\label{other-comps-1}
g^{(2)}_{uu} = - \frac{1}{2} R_0, ~~ g^{(2)}_{ua} =  \frac{1}{2} D^b g^{(1)}_{ba}, ~~ g^{(2)}_{ab} = \frac{1}{4} g^{(1)}_{ac} g_{(0)}^{cd} g^{(1)}_{db}, ~~ g_{(0)}^{ab} \, g^{(1)}_{ab} =0
\eea
\bea
\label{other-comps-2}
g^{(1)}_{ab} &=& g^{(1,0)}_{ab} (z, \bar z) + u \, g^{(1,1)}_{ab} (z, \bar z)
\eea
Along with the differential conditions:
\bea
\label{bms-cwi-1}
D^b \partial_ug^{(1)}_{ba} + \frac{1}{2} \partial_a R_0  =0 
\eea
\bea
\label{bms-cwi-2}
D_a g_{ub}^{(2)} - D_b g_{ua}^{(2)} + \frac{1}{4} \left( g^{(1)}_{ac} \, g_{(0)}^{cd} \,\partial_u g^{(1)}_{db} - g^{(1)}_{bc} \, g_{(0)}^{cd} \, \partial_u g^{(1)}_{da} \right)  &=& 0 
\eea
When these differential conditions (\ref{bms-cwi-1}, \ref{bms-cwi-2}) are solved for $g^{(0)}_{ab}$ and $g^{(1)}_{ab}$ then the rest of the 4d metric components $(g^{(2)}_{uu}$, $g^{(2)}_{ua}$, $g^{(2)}_{ab}$) can be found using (\ref{other-comps-1}, \ref{other-comps-2}).
The $u$-dependent part of the equation (\ref{bms-cwi-2}) is satisfied identically when we use (\ref{bms-cwi-1}) and thus can be simplified to:
\bea
g_{(0)}^{cd} \, \left[ \nabla_a \nabla_c g_{bd}^{(1,0)} - \nabla_b \nabla_c g_{ad}^{(1,0)} + \frac{1}{2} \left(  g^{(1,0)}_{ac} \,  g^{(1,1)}_{db} - g^{(1,0)}_{bc} \,   g^{(1,1)}_{da} \right) \right]  &=& 0 \, .
\eea
%
%
%
\subsection{Complete set of solutions}
It turns out that the equations (\ref{bms-cwi-1}, \ref{bms-cwi-2}) can be solved completely which we turn to now. 
%
To solve these equations first we take the boundary metric $g^{(0)}_{ab}$ in the form 
\bea
g^{(0)}_{ab} = \Omega \left( \begin{array}{cc} f & \frac{1}{2} (1+ f \, \bar f) \\ \frac{1}{2} (1 + f \, \bar f) & \bar f \end{array} \right)
\eea
with determinant: $-\frac{\Omega^2}{4} (1- f \bar f)^2$ where ($f$, $\bar f$, $\Omega$) are arbitrary functions of $(z, \bar z)$. Then we further parametrise $f(z, \bar z)$ and $\bar f(z, \bar z)$ as follows in terms of two other independent functions $\zeta(z, \bar z)$ and $\bar \zeta(z, \bar z)$: 
\bea
\label{beltramis}
f(z, \bar z) = \frac{\partial_z \bar \zeta(z, \bar z)}{\partial_{\bar z} \bar \zeta (z, \bar z)}, ~~~ \bar f (z, \bar z) = \frac{\partial_{\bar z}  \zeta(z, \bar z)}{\partial_z  \zeta (z, \bar z)}
\eea
Notice that this parametrisation is not one-to-one: suppose $\zeta(z, \bar z)$ ($\bar \zeta (z, \bar z)$) gives rise to a given $f(z, \bar z)$ ($\bar f(z, \bar z)$) then so does $\psi (\zeta(z, \bar z))$ ($\bar \psi (\bar \zeta(z, \bar z))$). Then the solutions for $g^{(1)}_{ab}$ components can be given explicitly in terms of the functions $(f, \bar f, \Omega)$ and $\rho = \left(\partial \zeta \bar \partial \bar \zeta\right)^{-1/2}$ and their derivatives. 

To keep the expressions simple we drop all the coordinate dependences $(z, \bar z)$ of various functions and use $\partial_z f(z, \bar z) \rightarrow \partial f$, $ \partial_{\bar z} f(z, \bar z)\rightarrow \bar \partial f $ etc. We define:
{\small
\bea
\{\chi(z, \bar z), z\} &=& \frac{\partial^3\chi(z, \bar z)}{\partial\chi(z, \bar z)} - \frac{3}{2} \frac{(\partial^2 \chi)^2}{(\partial \chi)^2}, ~~ \{\chi(z, \bar z), \bar z\} = \frac{\bar \partial^3\chi}{\bar \partial\chi} - \frac{3}{2} \frac{(\bar \partial^2 \chi)^2}{(\bar \partial \chi)^2}, \cr && \cr
~~ && \tau_{ab} = \frac{\partial_a \partial_b \Omega}{\Omega} - \frac{3 \, \partial_a \Omega \, \partial_b \Omega}{2 \, \Omega^2} 
\eea
}
where $a \in \{z, \bar z\}$.
Then components $g^{(1,0)}_{zz}$ and $g^{(1,0)}_{\bar z \bar z}$ are given by:
{\small
\bea
\label{g10zz}
 g^{(1,0)}_{zz}/\sqrt{\Omega} &=& -  \frac{\rho}{(1-f \bar f)^2} \left[2 (1+f^2 \bar f^2) \partial^2 \upsilon + 4 f^2 \, {\bar\partial}^2 \upsilon \right] \cr
&& + \partial \upsilon \Big[ \frac{2 \, \rho}{(1-f\bar f)^3} \left( - \bar f (1+ f^2 \bar f^2) \,\partial  f - (1-2f \bar f - f^2 \bar f^2) \,{\bar \partial} f - f (1+ f^2 \bar f^2) \,\partial{\bar f} + 2 f^2 \,{\bar \partial} \bar f \right) \cr
 && \hskip 3cm - \frac{4}{(1-f\bar f)^2} \left( (1+f^2 \bar f^2) \partial \rho - f (1+f \bar f) \,{\bar \partial} \rho \right) \Big] \cr
 && + \,{\bar \partial} \upsilon \, \Big[ \frac{2 \rho}{(1- f\bar f)^3} \left( (1+ f^2 \bar f^2) \, \partial f - 2 f^2 \bar f \,{\bar \partial} f + 2 f^3 \bar f \partial {\bar f} - 2 f^3 \,{\bar \partial} \bar f \right) \cr
 && \hskip 3cm + \frac{4f}{(1- f\bar f)^2} \left( (1+ f \bar f) \, \partial \rho - 2 f \,{\bar \partial} \rho \right) \Big]  + \frac{4 f (1+ f\bar f)}{(1-f \bar f)^2} \rho \, \partial {\bar \partial} \upsilon \cr && 
\eea
}
%
%
where $\upsilon = \upsilon(z, \bar z)$ is also arbitrary. Next, the $g^{(1,1)}_{zz}$ is given by
%
%
%
{\small 
\bea
\label{g11zz}
&& g^{(1,1)}_{zz} = \cr
&& \frac{\partial \Omega}{\Omega(1- f \bar f)^3} \Big[ - 2 f^2 \, {\bar \partial} \bar f + (1+ f^2 \bar f^2) f \, \partial {\bar f} + (1+ f^2 \bar f^2) \bar f \, \partial f + (1-2 f \bar f - f^2 \bar f^2)\, {\bar \partial} f \Big] \cr
&& + \frac{{\bar \partial} \Omega}{\Omega(1- f \bar f)^3} \Big[ -2 \bar f \, f^3 \, \partial {\bar f} + 2 f^3 \, {\bar \partial} \bar f + 2 f^2 \bar f \, {\bar \partial} f- (1+f^2 \bar f^2) \, \partial f \Big] \cr
&& + \frac{1}{(1-f\bar f)^2} \Big[ (1+ f^2 \bar f^2) \, \tau_{zz}  - 2 f \, (1+f \bar f) \tau_{z\bar z} + 2f^2 \tau_{\bar z \bar z} \Big] \cr
&& - \frac{1}{(1-f\bar f)^2} \Big[ (1- f \, \bar f)^2 \{\zeta, z \} 
+ f^2 \,  (1-f\bar f)^2 \,  \{\bar \zeta, \bar z \}\Big]  \cr
&& + \frac{1}{(1-f\bar f)^2} \Big[  2f \,  \partial^2 {\bar f} -2 f^2 \, \partial {\bar \partial} {\bar f} + f (1+ 2 f \bar f- f^2 \bar f^2) \,  {\bar \partial}^2 f  -(1+f^2 \bar f^2) \,  \partial  {\bar \partial} f \Big] \cr
&& + \frac{1}{(1-f\bar f)^3} \Big[ -2 f^3 \, \partial {\bar f} {\bar \partial} {\bar f} + f^2 \, (1+f \bar f) \,  (\partial {\bar f})^2 -\frac{1}{2} (1-3f \bar f -3 f^2 \bar f^2 + f^3 \bar f^3) \, ({\bar \partial} f)^2  \cr
&& \hskip 2.5cm  -\bar f (1+ f^2 \bar f^2) \,  \partial f \, {\bar \partial} f  + (1+f^2 \bar f^2) \,  \partial f \, \partial {\bar f}  -2f \, (1+f \bar f) \,  {\bar \partial} f \, \partial {\bar f} + 2f^2 \,  {\bar \partial} f \, {\bar \partial} {\bar f} \Big] \cr && 
\eea
}
The components $g^{(1,0)}_{\bar z \bar z}$ and $g^{(1,1)}_{\bar z \bar z}$ can be obtained from $g^{(1,0)}_{ z z}$ and $g^{(1,1)}_{z z}$ respectively by $f \leftrightarrow \bar f$ and $\partial \rightarrow \bar \partial$ (with $\rho$, $\upsilon$ and $\Omega$ unchanged).\footnote{As commented above under $(\zeta , \bar \zeta) \rightarrow (\psi(\zeta), \bar \psi(\bar \zeta))$ the functions $(f, \bar f)$ in (\ref{beltramis}) remain unaffected, but we have
\bea
\{ \zeta, z\} \rightarrow (\partial_z \zeta)^2 \, \{ \psi(\zeta), \zeta \} + \{\zeta, z\}, ~~~
\{\bar \zeta, \bar z \} \rightarrow (\partial_{\bar z} \bar \zeta)^2 \, \{ \bar \psi(\bar \zeta), \bar \zeta\} + \{\bar \zeta, \bar z \}
\eea
Under this change the resulting configurations remain LF solutions.} 

To summarise, a general locally flat metric solving equations (\ref{bms-cwi-1}, \ref{bms-cwi-2}) is parametrised by the following arbitrary set of functions:  $(\Omega(z,\bar z), \zeta(z, \bar z), \bar \zeta(z, \bar z))$ in  $g^{(0)}_{ab}$ and $\rho$, and $\upsilon(z, \bar z)$ in $g^{(1)}_{ab}$.\footnote{We have generated this solution by starting with flat spacetime with metric $ds^2_{{\mathbb R}^{1,3}} = - 2 \, du \, dr + r^2 \, dz \, d\bar z$ and making a finite coordinate transformation that keeps us in the chosen gauge. A similar exercise was done in \cite{Compere:2016jwb, compere:2018fr} with a different parametrisation of the boundary metric. We find our parametrisation better suited for the problem at hand. The solution in \cite{compere:2018fr} matches with our LF solution after the following identifications, $\Phi=\frac{\log{ \rho(z,\bar{z})^{-2}}}{\Omega(z,\bar{z})} $ and $G=\big(\zeta(z,\bar{z}),\bar{\zeta}(z,\bar{z})\big).$ }
%
%
The remaining components $g_{z \bar z}^{(1,1)}$ and $g_{z\bar z}^{(1,0)}$ can be obtained from the above ones using the tracelessness condition of $g^{(1)}_{ab}$:
\bea
g^{(1)}_{z\bar z} = \frac{1}{1+f \bar f} \Big[ g^{(1)}_{zz} \, \bar f + g^{(1)}_{\bar z \bar z} \, f \Big] \, .
\eea
\subsection{Gauge fixing the boundary metric}
\label{pg-lf-solns}
As shown in the appendix \ref{appendixA} the residual diffeomorphisms that preserve the NU gauge and the boundary conditions act as diffeomorphisms and Weyl transformations on $g^{(0)}_{ab}$.  These boundary diffeomorphisms can be used to gauge fix the $2d$ metric $g^{(0)}_{ab}$ down to one independent function. The most commonly used gauge choice for 2-dimensional metrics is the conformal gauge. In our language this amounts to setting $f(z, \bar z) = \bar f (z, \bar z) =0$ and letting $\Omega$ fluctuate. However, as we are interested in getting a symmetry algebra that includes $\mathfrak{sl}_2$ current algebra we choose the analogue of 
\bea
{\rm Polyakov ~ Gauge:} ~~   \bar f(z, \bar z)=0 ~{\rm or}~ f(z, \bar z) =0 ~~{\rm with}~~ \Omega ~~ {\rm fixed ~to ~a~ given~ function}.
\eea
%
Below we provide the solutions in this gauge.
%
\subsubsection{Locally flat solutions in Polyakov gauge}
To be specific, we choose $\bar f =0$. Then the components of $g^{(0)}_{ab}$ reduce to
\bea
g^{(0)}_{zz} = \Omega (z, \bar z) \, \frac{\partial_z \bar \zeta_(z, \bar z)}{\partial_{\bar z} \bar \zeta_(z, \bar z)} := \Omega(z, \bar z) \, f(z, \bar z), ~~ g^{(0)}_{z \bar z} = \frac{1}{2}\Omega(z, \bar z), ~~ g^{(0)}_{\bar z \bar z} =0
\eea
Similarly the components $g^{(1,0)}_{ab}$ and $g^{(1,1)}_{ab}$ in (\ref{g10zz}, \ref{g11zz}) reduce to:
\bea
\label{varpconds}
g^{(1,1)}_{\bar z \bar z} &=&  -\{ \bar \zeta(z, \bar z), \bar z \} + \tau_{\bar z\bar z}(\Omega) \cr && \cr
g^{(1,0)}_{\bar z \bar z} /\sqrt{\Omega}
&=& -2 \sqrt{\frac{\partial_{\bar z} \bar \zeta (z, \bar z)}{\partial_z \zeta(z)}} \, ~ \partial_{\bar z} \left[ \frac{\partial_{\bar z} \upsilon  (z, \bar z)}{\partial_{\bar z} \bar \zeta (z, \bar z)} \right]
\eea
%
%
\bea
g_{zz}^{(1,1)} &=& f^2 \, g_{\bar z \bar z}^{(1,1)} -  \{ \zeta(z), z \} + f \, \partial_{\bar z}^2 f - \frac{1}{2} \left( \partial_{\bar z} f \right)^2 - \partial_z \partial_{\bar z} f \cr && \cr
&& + \frac{\partial_z \Omega (z, \bar z)}{\Omega (z, \bar z)} \partial_{\bar z} f(z, \bar z) - \frac{\partial_{\bar z} \Omega (z, \bar z)}{\Omega (z, \bar z)} \partial_z f(z, \bar z) \cr && + \tau_{zz}(\Omega) \,+ f(z, \bar z)^2 \tau_{\bar z\bar z}(\Omega) - 2 \, f(z, \bar z) \, \tau_{z\bar z}(\Omega)
\eea
%
%
%
\bea
&& ~~~~ g_{zz}^{(1,0)}/ \sqrt{\Omega (z, \bar z)} = -2\, \rho \,   \left[ \partial_z^2 \upsilon + 2 \, f^2 \, \partial_{\bar z}^2 \upsilon -2 \, f \, \partial_z \partial_{\bar z} \upsilon \right]  \cr
&& + 2 \, \rho \, \frac{\partial_z^2 \zeta}{\partial_z \zeta} \, \left[ \partial_z \upsilon - 2 \, f \, \partial_{\bar z} \upsilon \right] + 2  \, \left[\rho \, \left( \partial_z f - 2 \, f \, \partial_{\bar z} f \right) - 2 \, f \, \partial_z \rho \right] \partial_{\bar z} \upsilon
\eea
where $\rho = \frac{1}{\sqrt{\partial_z \zeta(z) \, \partial_{\bar z} \bar \zeta(z,\bar z)}}$. We also have to take
\bea
g_{z \bar z}^{(1)} =  f(z, \bar z) \, g_{\bar z \bar z}^{(1)}
\eea
Now that we know $g^{(0)}_{ab}$ and $g^{(1)}_{ab}$ the rest of the components can be found using the relations (\ref{other-comps-1}, \ref{other-comps-2}). 
We will not be interested in general conformal factor $\Omega (z, \bar z)$ in what follows -- but only two choices $\Omega =1$ and $\Omega = 4 \, (1+ z \bar z)^{-2}$. These choices have the property that the quantities $\tau_{zz}, \, \tau_{\bar z \bar z}$ vanish.
%
%
%
\subsubsection{Restriction to $g^{(1)}_{\bar z\bar z} =0$}
\label{3.4.1}
In the next section we will show that the solutions with vanishing $g^{(1,1)}_{\bar z \bar z}$ and $g^{(1,0)}_{\bar z \bar z}$ also satisfy the variational principle. Anticipating that result we write down the solutions with $g^{(1)}_{\bar z\bar z} =0$. In the cases for $\Omega$'s with $\tau_{zz} = \tau_{\bar z \bar z} =0$, we see from (\ref{varpconds}) that 
\bea
\bar \zeta (z, \bar z) = \frac{g_1(z) \, \bar z + g_2(z)}{g_3(z) \, \bar z + g_4(z)}, ~~ \upsilon(z, \bar z) = \upsilon_{(1)}(z)  + \bar \zeta (z, \bar z) \, \upsilon_{(2)}(z)
\eea
%
%
%
Then $f(z, \bar z)=\frac{\partial_z \bar \zeta (z, \bar z)}{\partial_{\bar z} \bar \zeta (z, \bar z)}$  turns out to be a polynomial of degree two in $\bar z$:
\bea
\label{fzz}
(\det \mathfrak{g}) \,  f(z, \bar z) &=& \left[ g_2'(z) g_4(z) - g_2(z) g_4'(z) \right] \cr
&& + \left[ g_1'(z) g_4(z) - g_1(z) g_4'(z) +g_2'(z) g_3(z) - g_2(z) g_3'(z) \right] \, \bar z \cr
&& + \left[ g_1'(z) g_3(z) - g_1(z) g_3'(z) \right] \, \bar z^2 \cr &:=& - (\det \mathfrak{g}) \,\sum_{a=0}^{2} J_{a-1} (z) \, \bar z^{a}
\eea
where $\det \mathfrak{g} := g_1(z) \, g_4(z) - g_2(z) \, g_3(z)$. For (\ref{fzz}) to make sense we have to assume that $\det \mathfrak{g} \ne 0$.\footnote{There is another interesting interpretation of the coefficients $J_a$ of the powers of $\bar z$ in (\ref{fzz}) as follows. First consider the $2 \times 2$ matrix 
\bea
\mathfrak{g} = \left( \begin{array}{cc} g_1(z) & g_2(z) \\ g_3(z) & g_4(z) \end{array} \right)
\eea
with $\det \mathfrak{g} := g_1(z) \, g_4(z) - g_2(z) \, g_3(z) \ne 0$ making $\mathfrak{g} \in GL_2({\mathbb C})$. Then define $A(z) = \mathfrak{g}^{-1}(z)  \,\partial_z \mathfrak{g}(z)$ which is an element of the algebra $\mathfrak{gl}_2$. Consider a $2\times 2$ matrix representation of $sl_2$ algebra $[t^a, t^b] = (a-b) \, t^{a+b}$ -- for instance, in terms of the Pauli matrices we can take $t^{(0)} = \frac{1}{2} \sigma_3$, $t^{(\pm 1)}= \frac{i}{2} (\sigma_2 \pm i \, \sigma_1)$.  Then it turns out that 
\bea
J_a(z) = 2 \, \eta_{ab} \, {\rm Tr}  \left[ t^b \, A(z) \right] := \eta_{ab} \, J^b(z)
\eea
where the non-zero components of $\eta_{ij}$ are $\eta_{+1\, -1} = \eta_{-1\, +1} = 1/2$ and $\eta_{00} =-1$. One also has ${\rm Tr} \left[ A(z) \right] = \frac{1}{2 \, \det \mathfrak{g}} \partial_z \det \mathfrak{g}$ and without loss of generality we may choose $\det \mathfrak{g} =1$. This makes $A(z) \in \mathfrak{sl}_2$ and $A(z) = J_a(z) \, t^a$.}
Furthermore, 
\bea
g_{zz}^{(1,0)} &=& - 2 \, \sqrt{\frac{\Omega}{\zeta'(z) \partial_{\bar z} \bar \zeta (z, \bar z)}} \left[ \upsilon_{(1)}''(z) + \bar \zeta (z, \bar z) \, \upsilon_{(2)}''(z) - \frac{\zeta''(z)}{\zeta'(z)} \left(\upsilon_{(1)}'(z) + \bar \zeta (z, \bar z) \, \upsilon_{(2)}'(z) \right) \right]  \cr && 
\eea
Since $\partial_{\bar z} \bar \zeta (z, \bar z) = (\det \mathfrak{g}) /(g_3(z) \, \bar z + g_4(z))^2$ we see that  $g_{zz}^{(1,0)}/\sqrt{\Omega}$ is linear in $\bar z$. Therefore we write $g_{zz}^{(1,0)}$ as 
\bea
g_{zz}^{(1,0)} = \sqrt{\Omega} \, \left[ C_{-1/2} (z) + \bar z \, C_{1/2}(z) \right] 
\eea
where 
\bea
\label{crs}
&& C_{-1/2} (z) = \cr
&& ~~~~ - \frac{2}{ \det \mathfrak{g} \,  \sqrt{ \zeta'(z)}} \left[ g_4 (z) \left( \upsilon_{(1)}''(z) - \frac{ \zeta''(z)}{\zeta'(z)} \upsilon_{(1)}'(z) \right) + g_2 (z) \left( \upsilon_{(2)}''(z) - \frac{ \zeta''(z)}{\zeta'(z)} \upsilon_{(2)}'(z) \right) \right] \cr
&& C_{1/2} (z) = \cr
&& ~~~~ - \frac{2}{\det \mathfrak{g} \,  \sqrt{ \zeta'(z)}} \left[ g_3 (z) \left( \upsilon_{(1)}''(z) - \frac{ \zeta''(z)}{\zeta'(z)} \upsilon_{(1)}'(z) \right) + g_1 (z) \left( \upsilon_{(2)}''(z) - \frac{ \zeta''(z)}{\zeta'(z)} \upsilon_{(2)}'(z) \right) \right]  \cr &&
\eea
We can equivalently write $g_{zz}^{(1,0)}(z, \bar z) = \sqrt{\Omega (z, \bar z)} \, \sum_{r,s \in \{-1/2,1/2 \}}  \epsilon_{rs} \bar z^{r+1/2} \, C^{s}$ with $\epsilon_{-\frac{1}{2} \frac{1}{2}} =1$ with $C_{r}(z) = \epsilon_{rs} \, C^{s}(z)$.
%
Finally we have
\bea
g^{(1,1)}_{zz} &=& - \{ \zeta(z), z \} + \frac{1}{2} \, \eta_{ij} J^{i}(z) \, J^{j}(z) - \, \partial_z J^{(0)}(z) + \, \bar z \, \partial_z J^{(-1)}(z)  \cr
&& + \frac{\partial_z \Omega (z, \bar z)}{\Omega (z, \bar z)} \partial_{\bar z} f(z, \bar z) - \frac{\partial_{\bar z} \Omega (z, \bar z)}{\Omega (z, \bar z)} \, \partial_z f(z, \bar z)  -2 \, f(z, \bar z) \, \tau_{z\bar z} (\Omega)
\eea
%
%
%
%
Now we will leave the Polyakov gauge solutions\footnote{Given these Polyakov gauge solution we can readily find the solutions in conformal gauge. We simply have to set $f(z, \bar z) =0$ which can be done by choosing $\bar \zeta (z, \bar z) = \bar \zeta (\bar z)$.} found in this section in store until we learn to impose the variational problem -- which will be done in the next section.

\section{Boundary terms and variational principle}
\label{sec4}
We now turn to the variational problem among the configurations we have considered so far. To be precise we consider our configuration space to be all the 4-dimensional metrics $g_{\mu\nu}$ that are in the NU gauge which asymptotically approach the locally flat solutions. Then we will define our solution space to be a subset of Ricci flat configurations, and equally importantly, that also satisfy a variational principle $\delta S =0$. 

The usual prescription of boundary terms consists of adding a Gibbons-Hawking like term and a set of possible counter terms to the standard Einstein Hilbert action $S_{EH}$. There are two essential aspects required of such boundary terms: 
\begin{enumerate}
\item The boundary action $S_{bdy}$ has to be consistent with the symmetries that leave the gauge choice and the boundary surface invariant. 
\item The variation of $S_{EH} + S_{bdy}$ should be proportional to variations of metric data on the boundary - but not its derivatives. 
\end{enumerate}
The standard Gibbons-Hawking term is actually invariant under the full set of 3$d$ diffeomorphisms of the boundary. In the context of $AdS_{n+1}$ gravity in the Fefferman-Graham gauge residual symmetries are the diffeomorphisms of the $n$-dimensional subspace $r=r_0$ and thus adding the Gibbons-Hawking term and other counter terms that are also invariant under the $n$-dimensional diffeos is justified. 

Now we pose this question in our context: what are the subset of asymptotic symmetries of the class of geometries that leave $r=r_0$ fixed. Generators of any such coordinate transformation has to have $\xi^r =0$. These vectors of course have to continue to solve (\ref{asv1}, \ref{asv2}, \ref{asv3}). The first consequence of this condition $\partial_r \xi^r=0$, is that $(\partial_u - g_{ua} g^{ab} \partial_b) \, \xi^u =0$. Since $\xi^r_{(0)} = \xi^u_{(1)} = \partial_u \xi^u$ we also have $\partial_u \xi^u =0$ implying $g_{ua} g^{ab} \partial_b \, \xi^u =0$. From (\ref{asv3}) working in the gauge $g^{(0)}_{ab} g_{(1)}^{ab} =0$ we see that $\xi^r_{(1)}=0$ requires $\xi^u$ to be a harmonic function in 2$d$. Since the harmonic equation is Weyl invariant and in $2d$ every metric is Weyl equivalent to flat space the solutions to $\square \xi^u =0$ are $\xi^u(z, \bar z) = \xi^u_{(z)}(z) + \xi^u_{(\bar z)} (\bar z)$. Substituting this expression of $\xi^u$ into ${g_u}^a \partial_a \, \xi^u =0$ implies ${g_u}^z \, \partial_z \xi^u_{(z)} (z) + {g_u}^{\bar z} \, \partial_{\bar z} \xi^u_{(\bar z)} (\bar z) =0$. This equation is background dependent and a linear combination of a holomorphic and anti-holomorphic functions $(\partial_z \xi^u_{(z)} (z),\partial_{\bar z} \xi^u_{(\bar z)} (\bar z))$ and unless the coefficients $({g_u}^z, {g_u}^{\bar z})$ are zero (and non-generic) the only solutions are $(\partial_z \xi^u_{(z)} (z) =0 , \partial_{\bar z} \xi^u_{(\bar z)} (\bar z)=0)$ which in turn implies $\partial_a \xi^u=0$ and this is what we work with. To summarise the subset of symmetries of our class of geometries that leave $r=r_0$ invariant are: $(\xi^r=0, \xi^u = \xi^u_{(0)}, \xi^a = \xi^a_{(0)}(z, \bar z))$ -- that is, rigid translations in $u$ and arbitrary diffeomorphisms of $(z, \bar z)$. 

Another way to arrive at the same conclusion is the following. We should have expected that the residual transformations do not mix different orders of powers of $r$. This in turn implies that $\xi^a$ is independent of $r$ ($\xi^u$ is already independent of $r$). Then the last of (\ref{asv1}) implies $\partial_a \xi^u$ is zero, and the fact that $\xi^r$ should also vanish requires $\partial_u \xi^u=0$, thus making $\xi^u$ a constant and $\xi^a = \xi^a_{(0)} (z, \bar z)$. These are the generators of the boundary coordinate transformations: $(u \rightarrow u' = u+u_0, x^a \rightarrow {x'}^a = x'^a (x))$. The Jacobian of such transformations is: $\frac{\partial u'}{\partial u} =1$, $\frac{\partial u'}{\partial x^a} = \frac{\partial {x'}^a}{\partial u} =0$ with the only non-trivial part $J^a_b = \frac{\partial {x'}^a}{\partial {x}^b}$. Now we seek the boundary terms that respect at least these symmetries. 

The bulk four-dimensional metrics in the NU gauge are of the form:
\bea
g_{\mu\nu} = \left( \begin{array}{ccc} g_{uu}~~ & -1 & ~~~g_{ua} \\ -1~~ & 0 & ~~~0 \\ g_{au}~~ & 0 & ~~~g_{ab} \end{array} \right), ~~~ g^{\mu\nu} = \left( \begin{array}{rcl} 0 & ~~ -1 ~~ & 0 \\ -1 & ~~ g_{ua} {g_u}^a \! - \! g_{uu} ~~ & {g_u}^b \\ 0 & {g^a}_u & g^{ab} \end{array} \right)
\eea
where $g^{ab}$ is the inverse of $g_{ab}$, ${g_u}^a = g_{ub} g^{ba}$, etc. We take the unit normal to the $r=r_0$ surface as: $n_\mu = \frac{1}{\sqrt{g_{ua} {g_u}^a - g_{uu}}} \delta^r_\mu$ and $n^\mu =\frac{1}{\sqrt{g_{ua} {g_u}^a - g_{uu}}}g^{r\mu}$ such that $n_\mu n^\mu = 1$. 
%
%
The induced metric on $r=r_0$ surface is:
\bea
\gamma_{ij} = g_{\mu\nu} \, e^\mu_i e^\nu_j =\left( \begin{array}{cc} g_{uu}~~ & g_{ua} \\  g_{au}~~ & g_{ab} \end{array} \right)
\eea
with its inverse
\bea
\gamma^{ij} = \frac{1}{N^2} \left( \begin{array}{cc} -1 &~~~~ g^a_u \\  g^a_u &~~~~ N^2 \, g^{ab} - g^a_u g^b_u \end{array} \right)
\eea
with $N^2 = - g_{uu} + g_{au} \, g^a_u$ and the usual completeness relation $g^{\mu\nu} = n^\mu n^\nu + \gamma^{ij} e^\mu_i e^\nu_j$. Thus the metric data in the line element 
\bea
ds^2_{bdy} &=& g_{uu} \, du^2 + 2 \, g_{ua} du \, dx^a + g_{ab} \, dx^a \, dx^b
\eea
%
%
on the boundary is $(g_{uu}, g_{ua}, g_{ab})$. The expected set of symmetries of the boundary $(u \rightarrow u' = u+u_0, x^a \rightarrow {x'}^a = x'^a (x))$ is much smaller than the $3d$ diffeos in $(u, x^a)$. Therefore, we expect much more freedom in the choice of the boundary terms. The transformations of the components of the induced metric under the reduced symmetry are: 
\bea
\label{gunderrs}
g_{u'u'} (u', x'^a) = g_{uu} (u, x^a), ~~g_{u' x'^a} (u', x') = \frac{\partial x^b}{\partial x'^a} g_{u x^b} (u, x), ~~ g_{x'^a x'^b} (u', x') = \frac{\partial x^c}{\partial x'^a} \frac{\partial x^d}{\partial x'^a} g_{x^c x^d} (u, x). \cr && 
\eea
This enables us to classify the basic scalars under the boundary symmetries and some of them are:
\bea
{\rm no} ~ {\rm derivatives:} && g_{uu}, ~~ g^{ab} g_{ua} g_{ub}, \cr
{\rm one} ~ {\rm derivative:}&& \partial_u g_{uu}, g^{ab} g_{ua} \partial_u g_{ub}, ~ g_{ua} g_{ub} \partial_u g^{ab} ~ g^{ab} \partial_u g_{ab}, ~ D^a g_{ua}, \cr
{\rm two} ~ {\rm derivative:}&& R[g_{ab}], ~ \square g_{uu}, ~ \partial_u^2 g_{uu}, \partial_u D^a g_{ua}, ~ \partial_u g^{ab} \partial_u g_{ab}, ~ g^{ab} \, \partial_u g_{ua} \partial_u g_{ub} ~ etc.
\eea
The integration measure invariant under our boundary symmetry $(u, x^a) \rightarrow (u+u_0, x'^a(x))$ is $\int du \int d^2 x \, \sqrt{\sigma} $ where $\sigma = \det g_{ab}$, which can be used to integrate any function of the scalars listed above to obtain a potential boundary term. 
%
%
%
%
\subsection{The boundary terms}
Now we look for the potential Gibbons-Hawking type terms we can construct consistent with our boundary symmetries. The bulk action is the standard Einstein-Hilbert action 
\bea
S_{EH} = \frac{1}{16 \, \pi \, G} \int d^4x \, \sqrt{-g} \, R \label{EHterm}
\eea
Then the variation of the action within our configuration space around a configuration satisfying the Einstein equation $R_{\mu\nu} =0$ is 
\bea
\delta S_{EH} = \frac{1}{16 \, \pi \, G} \int d^4 x \, \sqrt{-g} \, g^{\mu\nu} \left( \nabla_\kappa \delta \Gamma^\kappa_{\mu\nu} - \nabla_\nu \, \delta \Gamma^\kappa_{\mu\kappa} \right) = \frac{1}{16 \, \pi \, G} \int_\Sigma d^3 x \, \sqrt{-\gamma} ~ n_\mu J^\mu 
\eea
where $J^\kappa = g^{\mu\nu} \, \delta \Gamma^\kappa_{\mu\nu} - g^{\mu\kappa} \, \delta \Gamma^\nu_{\mu\nu}$ where $\delta\Gamma^\kappa_{\mu\nu} = \frac{1}{2} g^{\kappa\omega} \, \left( \nabla_\mu \delta g_{\omega\nu} + \nabla_\nu \delta g_{\mu\omega} - \nabla_\omega \delta g_{\mu\nu} \right)$ which leads to $J^\mu =  g^{\mu\nu} g^{\sigma\lambda} \, \left( \nabla_\sigma \delta g_{\nu\lambda} - \nabla_{\nu} \delta g_{\sigma\lambda} \right)$. 
We will have to manipulate this term carefully. Since our surface is defined by $r=r_0$ the unit normal is $n_\mu = N \delta_\mu^r$ where $N = \frac{1}{\sqrt{g^{rr}}}$. But we also have $g^{rr} = \frac{\gamma}{g}$ where $\gamma = \det(\gamma_{ij})$ and $g = \det g_{\mu\nu}$ $\implies \sqrt{-\gamma} = N \, \sqrt{-g}$. 
%
%
%
This boundary term can be written explicitly for geometries in the NU gauge with $n_\mu = \frac{1}{\sqrt{g^{rr}}} \delta^r_\mu$ and we find:
\bea
\sqrt{-\gamma} \, n_\mu J^\mu &=& \sqrt{ \sigma} \, \partial_r \delta \omega + \frac{1}{2} \sqrt{\sigma} \, \sigma^{ab} \partial_r \sigma_{ab} \, \delta \omega + \frac{1}{2} \sqrt{\sigma} \, \omega \, \partial_r \sigma^{ab} \delta \sigma_{ab}  + \sqrt{\sigma} \, \omega \, \sigma^{ab} \partial_r \delta \sigma_{ab} \cr
&& - \sqrt{\sigma} \, v_a \partial_r v_b \, \delta \sigma^{ab} + \sqrt{\sigma} \, v^a v^b \, \partial_r \delta \sigma_{ab} + \partial_r(\sqrt{\sigma}) \, v^a v^b  \delta \sigma_{ab} + \sqrt{\sigma} \, v^a v_b \, \partial_r \sigma^{cb} \, \delta \sigma_{ac} \cr
&& - \frac{1}{2} \sqrt{\sigma} \, v_a v^a \, \partial_r \sigma_{ab} \delta \sigma^{ab} - \sqrt{\sigma} \, v^c v_c \, \sigma^{ab} \, \partial_r \delta \sigma_{ab} \cr
&& - \sqrt{\sigma} \, \sigma^{ab} \, \partial_r v_a \, \delta v_b - \sqrt{\sigma} \, v_a \partial_r \sigma^{ab} \, \delta v_b - 2 \, \partial_r (\sqrt{\sigma}) \, v^a  \, \delta v_a - 2 \, \sqrt{\sigma} \, v^a \partial_r \delta v_a \cr
&& + \frac{1}{2} \sqrt{\sigma} \, \partial_u \sigma_{ab} \, \delta \sigma^{ab} + \sqrt{\sigma} \, \sigma^{ab} \, \partial_u \delta \sigma_{ab} \cr
&& + \sqrt{\sigma} \left( D_a \delta \sigma_{bc} \right) \, \left( \sigma^{ab} \, v^c - v^a \,\sigma^{bc} \right)  - \partial_a \left( \sqrt{\sigma} \, \sigma^{ab} \delta v_b \right) 
\eea
where we introduced the notation: $\omega = g_{uu}$, $v_a = g_{ua}$ and $\sigma_{ab} = g_{ab}$ along with $\sigma^{ab}$ being the inverse of $\sigma_{ab}$ and $v^a = \sigma^{ab} v_b$. In this we seek that those terms with tangential derivatives (that is, derivatives w.r.t $(u, z, \bar z)$) on variations of the boundary data $(\delta \omega, \delta v_a, \delta \sigma_{ab})$ have to be cancelled. The usual strategy involves completing such terms into either total variations or total derivatives so that the variations (derivatives) are moved away from the derivatives (variations). To do this one may use the following identities:
\bea
\sqrt{ \sigma} \, \partial_r \delta \omega = \delta \left( \sqrt{\sigma} \, \partial_r \omega \right) - \partial_r \omega \, \delta \left(\sqrt{\sigma} \right) \nonumber
\eea
\bea
&& \hskip 3cm \sqrt{\sigma} \, \left (\omega \, \sigma^{ab} +  v^a v^b - v^c v_c \, \sigma^{ab} \right) \, \partial_r \delta \sigma_{ab} \cr && \cr
&&= \delta \, \left[ \sqrt{\sigma} \, \left (\omega \, \sigma^{ab} +  v^a v^b - v^c v_c \, \sigma^{ab} \right) \, \partial_r \sigma_{ab}\right] - \partial_r \sigma_{ab} \,\,  \delta \left( \sqrt{\sigma} \, \left (\omega \, \sigma^{ab} +  v^a v^b - v^c v_c \, \sigma^{ab} \right) \right)  \nonumber
\eea
\bea
- 2 \, \sqrt{\sigma} \, v^a \partial_r \delta v_a= \delta \left[- 2 \, \sqrt{\sigma} \, v^a \partial_r  v_a \right]+ 2 \, \partial_r  v_a \, \delta \left(\sqrt{\sigma} \, v^a \right) \nonumber
\eea
\bea
\sqrt{\sigma} \, \sigma^{ab} \, \partial_u \delta \sigma_{ab} =
\delta \left[ \sqrt{\sigma} \, \sigma^{ab} \, \partial_u  \sigma_{ab}\right] - \partial_u  \sigma_{ab} \, \delta\left[\sqrt{\sigma} \, \sigma^{ab}\right] \nonumber
\eea
\bea
&& \hskip 3cm \sqrt{\sigma} \left( D_a \delta \sigma_{bc} \right) \, \left( \sigma^{ab} \, v^c - v^a \,\sigma^{bc} \right) \cr
&& = \partial_a \left[\sqrt{\sigma} \, \left( \sigma^{ab} \, v^c - v^a \,\sigma^{bc} \right) \, \delta \sigma_{bc} \right] - \sqrt{\sigma} \, \delta \sigma_{bc} \, \left[ \sigma^{ab} \, D_a v^c - \sigma^{bc} D_a v^a  \right] \nonumber
\eea
Using these and moving the total variations and total derivatives (in $(z, \bar z)$ coordinates) to the lhs we find:
\bea
&& \sqrt{-\gamma} \, n_\mu J^\mu - \delta \left( \sqrt{\sigma} \, \partial_r \omega \right) - \delta \, \left[ \sqrt{\sigma} \, \left (\omega \, \sigma^{ab} +  v^a v^b - v^c v_c \, \sigma^{ab} \right) \, \partial_r \sigma_{ab}\right] \cr
&& + \delta \left[2 \, \sqrt{\sigma} \, v^a \partial_r  v_a \right] - \delta \left[ \sqrt{\sigma} \, \sigma^{ab} \, \partial_u  \sigma_{ab}\right]
 - \partial_a \left[\sqrt{\sigma} \, \left( \sigma^{ab} \, v^c - v^a \,\sigma^{bc} \right) \, \delta \sigma_{bc} \right] + \partial_a \left( \sqrt{\sigma} \, \sigma^{ab} \delta v_b \right)\cr && \cr
&& \hskip .5cm = - \partial_r \omega \, \delta \left[\sqrt{\sigma} \right]  
- \partial_r \sigma_{ab} \,\,  \delta \left[ \sqrt{\sigma} \, \left (\omega \, \sigma^{ab} +  v^a v^b - v^c v_c \, \sigma^{ab} \right) \right] \cr
&& \hskip 1cm + \frac{1}{2} \sqrt{\sigma} \, \sigma^{ab} \partial_r \sigma_{ab} \, \delta \omega + \frac{1}{2} \sqrt{\sigma} \, \omega \, \partial_r \sigma^{ab} \delta \sigma_{ab}  
\cr
&& \hskip 1cm - \sqrt{\sigma} \, v_a \partial_r v_b \, \delta \sigma^{ab} 
+ \partial_r(\sqrt{\sigma}) \, v^a v^b  \delta \sigma_{ab} 
+ \sqrt{\sigma} \, v^a v_b \, \partial_r \sigma^{cb} \, \delta \sigma_{ac} 
- \frac{1}{2} \sqrt{\sigma} \, v_c v^c \, \partial_r \sigma_{ab} \delta \sigma^{ab} \cr
&& \hskip 1cm - \sqrt{\sigma} \, \sigma^{ab} \, \partial_r v_a \, \delta v_b - \sqrt{\sigma} \, v_a \partial_r \sigma^{ab} \, \delta v_b - 2 \, \partial_r (\sqrt{\sigma}) \, v^a  \, \delta v_a + 2 \, \partial_r  v_a \, \delta \left[\sqrt{\sigma} \, v^a \right] \cr
&& \hskip 1cm + \frac{1}{2} \sqrt{\sigma} \, \partial_u \sigma_{ab} \, \delta \sigma^{ab} - \partial_u  \sigma_{ab} \, \delta\left[\sqrt{\sigma} \, \sigma^{ab}\right] \cr
&& \hskip 1cm - \sqrt{\sigma} \, \delta \sigma_{bc} \, \left[ \sigma^{ab} \, D_a v^c - \sigma^{bc} D_a v^a  \right] 
\eea
Note that there are no terms involving tangential derivative of $(\delta \omega, \delta v_a, \delta \sigma_{ab})$ on the rhs. Thus the boundary action required to be added to the Einstein-Hilbert action (\ref{EHterm}) in our context is,
\bea
(16 \pi G)\, S_{bdy.}&&= \int d^3 x \sqrt{\sigma} \left[ 2 \,  \, v^a \partial_r  v_a -  \partial_r \omega  - \left (\omega \, \sigma^{ab} +  v^a v^b - v^c v_c \, \sigma^{ab} \right) \, \partial_r \sigma_{ab}   -  \sigma^{ab} \, \partial_u  \sigma_{ab} \right] \cr
&&  = \int d^3 x\sqrt{\sigma} \, \partial_r  (v^a v_a - \omega) +2 \, \left[ (v^a v_a - \omega ) \,  \partial_r - \partial_u \right] \, \sqrt{\sigma} \label{bdryterm}
\eea
and the total divergence term in $\delta S_{EH}$ is $(16 \pi G)^{-1}$ times:
\bea
\partial_a \left[\sqrt{\sigma} \, \left( \delta v^a - v^a \, \sigma_{bc} \, \delta \sigma^{bc} \right) \right]
\eea
that we ignore (this amounts to assuming that the geometry of $\Sigma_2$ with coordinates $(z, \bar z)$ is either compact or, when it is not, the integrand falls-off fast enough near its asymptotes). 

In summary we have managed to find the boundary action (\ref{bdryterm}) we sought such that the variation of the total (bulk plus boundary) action is a linear combination of $(\delta \omega, \delta v_a, \delta \sigma_{ab})$ but not their derivatives: 
\bea
(16\pi G)\,\delta S_{EH+ bdy.} &=& \int_{r=r_0} du \, d^2 z \, \sqrt{\sigma} \, \left[ \mathfrak{t}_{ab} \, \delta \sigma^{ab} + \mathfrak{j}_a \, \delta v^a + \mathfrak{o} \, \delta \omega \right] \cr
&=& \int_{r=r_0} du \, d^2 z \, \sqrt{\sigma} \, \left[ - (\mathfrak{t}^{ab} + \mathfrak{j}^{(a} v^{b)}) \, \delta \sigma_{ab} + \mathfrak{j}^a \delta v_a + \mathfrak{o} \, \delta \omega \right] 
\eea
where $\mathfrak{t}^{ab} = \sigma^{ac} \sigma^{bd} \mathfrak{t}_{cd}$, $\mathfrak{j}^a = \sigma^{ab} \mathfrak{j}_b$. Finally we can expand the integrand $\delta {\cal L}_{EH +bdy}$ around large-$r$,
and we find:
{\small
\bea
&& \delta {\cal L}_{EH+bdy.} = -2 \, r^3 \, \sqrt{g^{(0)}} \, \delta g_{uu}^{(0)} - r^2 \sqrt{g^{(0)}} \, \left[ \frac{1}{2} \left(g_{(0)}^{cd} g^{(1)}_{cd} \right) \, \delta g_{uu}^{(0)} + 2 \, \delta g_{uu}^{(1)} \right] \cr
&& - r \, \sqrt{g^{(0)}} \, \Big[2 \, \delta g_{uu}^{(2)} + g_{uu}^{(2)} \, g_{(0)}^{ab} \, \delta g^{(0)}_{ab} + \frac{1}{2} \left(g_{(0)}^{cd} g^{(1)}_{cd} \right) \, \delta g_{uu}^{(1)} - 2 \, g^{(2)}_{ua} g_{(0)}^{ab} \delta g^{(0)}_{ub}  \cr
&& ~~~~~~~~~~~~~~~~ - \frac{1}{2} \left( g_{(0)}^{ab}\partial_u g^{(1)}_{bc} g_{(0)}^{cd} \, \delta g^{(0)}_{da} - \left(g_{(0)}^{cd} \partial_u g^{(1)}_{cd} \right) \left(g_{(0)}^{ab} \, \delta g^{(0)}_{ab} \right) \right]  - \sqrt{g^{(0)}} \,\,  \delta {\cal L}_0 + {\cal O}(1/r)  \cr &&
\eea
}
where:
\bea
&& \delta {\cal L}_0 = 2 \, \delta g^{(3)}_{uu} + \frac{1}{2} g^{(3)}_{uu} \, \left( g_{(0)}^{cd} \delta g^{(0)}_{cd} \right) + \frac{1}{2} \left(g_{(0)}^{cd}  g^{(1)}_{cd} \right) \, \delta g_{uu}^{(2)} + \frac{1}{2} g_{uu}^{(2)} \left(g_{(1)}^{ab} \delta g^{(0)}_{ab} \right) 
\cr
&& - \frac{1}{2} \, \Big[ \tr g^{(3)}  + \frac{1}{2} \tr g^{(1)} \, \tr g^{(2)}  - \tr \left( g^{(1)} g_{(0)}^{-1} g^{(2)} g_{(0)}^{-1} \right) \cr
&& - \frac{1}{8} \tr g^{(1)} \tr \left( g^{(1)} g_{(0)}^{-1} g^{(1)} g_{(0)}^{-1} \right) + \frac{1}{4} \tr \left( g^{(1)} g_{(0)}^{-1} \right)^3 \Big] \, \delta g_{uu}^{(0)} -  \left(D_{(a} g^{(2)}_{b)u}- D^c g_{uc}^{(2)} \, g^{(0)}_{ab} \right) \, \delta g_{(0)}^{ab}\cr
&& - \frac{1}{4} \tr g^{(1)} \, \left(\partial_u g^{(1)}_{ab} - \partial_u (\tr g^{(1)}) g^{(0)}_{ab} \right) \, \delta g_{(0)}^{ab} +\frac{1}{2} \left(  \partial_u g^{(2)}_{ab} \delta g_{(0)}^{ab} - \tr (\partial_u g^{(2)}) \, \left( g^{(0)}_{ab} \, \delta g_{(0)}^{ab} \right) \right) \cr
&& + g^{(2)}_{ua}  g_{(1)}^{ab}  \delta g^{(0)}_{bu} - g^{(3)}_{au} g_{(0)}^{ab} \, \delta g^{(0)}_{bu} - 2 g^{(2)}_{au} g_{(0)}^{ab} \, \delta g^{(1)}_{bu} + g^{(2)}_{uu} \, g_{(0)}^{ab} \delta g^{(1)}_{ab} \cr
&& - \frac{1}{2} \left( \tr \left( g_{(0)}^{-1} \partial_u g^{(1)} g_{(0)}^{-1} \delta g^{(1)}\right) - \tr \left( g_{(0)}^{-1} \partial_u g^{(1)} \right) \tr \left(g_{(0)}^{-1} \delta g^{(1)}\right)  \right)
\eea
with ${\rm Tr} g^{(2)} = g_{(0)}^{ab} g^{(2)}_{ab}$ and so on. As we will be taking the boundary $r=r_0 \rightarrow \infty$ we will not need to keep terms that vanish in this limit.
\subsection{Conditions from variational principle}
We first impose
\bea
\delta g^{(0)}_{uu} = \delta g^{(1)}_{uu} = \delta g_{ua}^{(0)} = \delta g_{ua}^{(1)} =0
\eea
for all configurations as these are part of the boundary conditions derived in Section \ref{sec3}. Furthermore, since we consider the class of geometries that are locally flat we restrict to the class (\ref{lf-configs}) of geometries: $ g^{(2+n)}_{\mu\nu} = \delta g^{(2+n)}_{\mu\nu} =0$ for $n=1,2,\cdots$. We also impose the gauge conditions and ${\rm Tr} \, g^{(1)} =0$ as well. Then we have
\bea
\label{finaldeltaS1}
\delta {\cal L} &=&   -2 \, r \, \delta \left(\sqrt{g^{(0)}} \, g_{uu}^{(2)} \right) + r \, \sqrt{g^{(0)}} \, \left[ \frac{1}{2} \, g_{(0)}^{ab} \partial_u g^{(1)}_{bc} g_{(0)}^{cd} \, \delta g^{(0)}_{da} \right]  - \sqrt{g^{(0)}} \, \, \delta {\cal L}_0 + {\cal O}(1/r) 
\eea
{\small 
\bea
\label{finaldeltaS2}
 \delta {\cal L}_0/\sqrt{g^{(0)}} &=&  \frac{1}{2} g_{uu}^{(2)} \left(g_{(1)}^{ab} \delta g^{(0)}_{ab} \right)  -  \left(D_{(a} g^{(2)}_{b)u}- D^c g_{uc}^{(2)} \, g^{(0)}_{ab} \right) \, \delta g_{(0)}^{ab} \cr
&& \!\!\!\!\!\!\!\!\!\!\!\! + \frac{1}{2} \left(  \partial_u g^{(2)}_{ab} \delta g_{(0)}^{ab} - \tr (\partial_u g^{(2)}) \, \left( g^{(0)}_{ab} \, \delta g_{(0)}^{ab} \right) \right) + g^{(2)}_{uu} \, g_{(0)}^{ab} \delta g^{(1)}_{ab} - \frac{1}{2} \left( \tr \left( g_{(0)}^{-1} \partial_u g^{(1)} g_{(0)}^{-1} \delta g^{(1)}\right)  \right) \cr &&
\eea
}
We just have to ensure that this quantity $\delta {\cal L}$, when integrated over the boundary directions $(u, z, \bar z)$, vanishes in the $r=r_0 \rightarrow \infty$. One more condition that we will impose for all the cases below is:
\bea
\delta \int d^2z  \left[\sqrt{g^{(0)}} \, g_{uu}^{(2)} \right] =0
\eea
which when using the relation $g_{uu}^{(2)} = - \frac{1}{2} R_0$ is equivalent to holding the Euler Character of the metric on $\Sigma_2$ fixed under variations (see also \cite{compere:2018fr}). Now we consider special cases of boundary conditions that ensure $\delta S =0$.

\subsubsection{Solutions with Dirichlet Boundary conditions}
In this case we take $\delta g^{(0)}_{ab} =0$. Since $g_{uu}^{(2)}$ is related to the curvature of $g^{(0)}_{ab}$ as given in (\ref{other-comps-1}) it follows that $\delta g_{uu}^{(2)} =0$ and therefore $\delta \left( \sqrt{g^{(0)}} \, g_{uu}^{(2)} \right) =0$. Using $\delta g^{(0)}_{ab} = 0$ in the rest of the terms leaves:
\bea
\delta {\cal L}_0 &=& - \sqrt{g^{(0)}} \, \left[g^{(2)}_{uu} \, g_{(0)}^{ab} \delta g^{(1)}_{ab} - \frac{1}{2} \left( \tr \left( g_{(0)}^{-1} \partial_u g^{(1)} g_{(0)}^{-1} \delta g^{(1)}\right)  \right) \right]
\eea
The first term in the bracket is proportional to the variation of $g_{(0)}^{ab} g^{(1)}_{ab}$ which is held to zero. So the only non-trivial term in the variation of the Lagrangian is proportional to $g_{(1,1)}^{ab} \delta g^{(1)}_{ab}$, and the simplest (covariant and non-chiral) way to ensure the vanishing of this term is to set $g^{(1,1)}_{ab} =0$. Thus we conclude that the locally flat solutions that satisfy the variational principle with fixed metric on the boundary $\Sigma_2$ is given by the class in (\ref{lf-cg}) with $\kappa (z) = \bar \kappa(\bar z) =0$ whenever $\Omega$ corresponds $\Sigma_2$ being ${\mathbb R}^2$ ($\Omega =1$) or $S^2$ ($\Omega = 4 \, (1+ z \bar z)^{-2})$ or ${\mathbb H}^2$ $(\Omega = 4 \, (1- z \bar z)^{-2})$.

There is a very interesting consequence of this conclusion: The algebra of symmetries that enable us to move along the space of locally flat solutions in this case is exactly $\mathfrak{bms}_4$, but do not allow for the $\mathfrak{bms}_4$ algebra to become the extended $\mathfrak{bms}_4$ of \cite{Bondi:1962bm, Sachs:1962sa, Sachs:1962rs}.
\subsection{Polyakov gauge solutions}
\label{varPolyakov}
Next we turn to the non-Dirichlet boundary metrics on $\Sigma_2$ and in particular restrict to the Chiral Polyakov gauge for the metric on $\Sigma_2$:  $\bar f(z, \bar z)=0$ with $\Omega$ to be given function of $(z, \bar z)$. We will continue to impose the boundary conditions 
\bea
\delta g_{uu}^{(0)} = \delta g_{uu}^{(1)} = \delta g_{ua}^{(0)} = \delta g_{ua}^{(1)} = \delta \int d^2z \sqrt{g^{(0)}} R_0 =0
\eea
In this gauge even though $\delta g_{zz}^{(0)} \ne 0$ we still have $\delta \sqrt{\det g^{(0)}_{ab}} = 0$.  We will need to ensure the remaining terms in (\ref{finaldeltaS1}, \ref{finaldeltaS2}) also vanish.
\begin{enumerate}
\item From the ${\cal O}(r)$ term in $\delta {\cal L}$  that is proportional to $g_{\bar z \bar z}^{(1,1)} \delta F$ we need to set $g_{\bar z\bar z}^{(1,1)} =0$. 
\item From the first term in ${\cal O}(r^0)$ terms we further need to impose $g_{\bar z \bar z}^{(1,0)} =0$.  
\item Then imposing the conditions $g_{(0)}^{ab} g^{(1)}_{ab} =0$ immediately implies $g_{z\bar z}^{(1)} =0$. 
\item Substituting $g_{z\bar z}^{(1)} = g_{\bar z\bar z}^{(1)} =0$ into $g^{(2)}_{ab} = \frac{1}{4} g^{(1)}_{ac} g_{(0)}^{cd} g^{(1)}_{db}$ we find that $g^{(2)}_{ab} =0$. 
\item Further in our gauge $g_{(0)}^{ab} \, \delta g^{(0)}_{ab} =0$, $g_{(0)}^{ab} \, \delta g^{(1)}_{ab} =0$, and ${\rm Tr}\left( g_{(0)}^{-1} \partial_u g^{(1)} g_{(0)}^{-1} \delta g^{(1)}\right) =0$. 
\item Finally we need to check: $D_{\bar z} g^{(2)}_{\bar z u} = D_{\bar z} D^a g^{(1)}_{a \bar z}$. Using  $g^{(1)}_{z\bar z} = g^{(1)}_{\bar z \bar z} =0$ and (\ref{other-comps-1}) it is easy to see that this quantity also vanishes.
\end{enumerate}
To summarise the variational principle in the Polyakov gauge can be solved by imposing:
\bea
g_{\bar z \bar z}^{(1,0)} = g_{\bar z \bar z}^{(1,1)} =0
\eea
on the class of locally flat geometries we found in the previous section. 

In anticipation of this result we have already presented the solutions satisfying these additional conditions in section \ref{3.4.1}. Here we present the locally flat solutions in closed form for two cases: $\Omega =1$ and $\Omega = 4 \, (1+ z \bar z)^{-2}$ that also are consistent with our variational principle.
\begin{itemize}
\item Locally flat solutions $\Omega =1$:
\begin{align}
\label{finalLFr2}
ds^2_{M_2 = {\mathbb R}^2}=& -\,2\,du\,dr +\,r^2\,dz\,d\bar{z} +r^2\,\big(-\eta_{ij}\,J^i\,\bar{z}^{1+j}\big)\,dz^2 \cr
&+r\,u\,\bigg(\,\frac{1}{u}\,\epsilon_{sr}\,C^r\,\bar{z}^{\frac{2s+1}{2}}-\kappa(z) +\frac{1}{2} \eta_{ij} J^i (z) J^j (z) +(1+i)\,\eta_{ij}\, \bar{z}^i\,\partial_zJ^j\bigg)\,dz^2 \cr
& +2 \,\left(\epsilon_{rs}\,\frac{2s+1}{2}\,C^r\,\bar{z}^{\frac{2s-1}{2}}+\,u\,\eta_{ij}\,j\,(1+j)\,\partial_z J^i\,\bar{z}^{j-1}\right)\,du\,dz \cr
& + 2\,\eta_{ij}\,j\,(1+j)\,J^a\,\bar{z}^{b-1}\,du^2 
\end{align}
%
%
\item Locally flat solutions with $\Omega = 4 \, (1+ z \bar z)^{-2}$:
\begin{align}
\label{finalLFs2}
ds^2_{\mathcal{S}^2}=&-2\,du\,dr+\frac{4\,r^2}{(1+z\,\bar{z})^2}\,dz\,d\bar{z} -\frac{4\,r^2}{(1+z\,\bar{z})^2}\, \eta_{ij}\,J^j\,\bar{z}^{1+b} \,dz^2\cr
&+r\,u\,\bigg[-\kappa(z)+\frac{1}{2} \eta_{ij} J^i J^j+(1+i)\,\eta_{ij}\, \bar{z}^i\,\partial_z J^j \cr
& \hskip 1.5cm +\frac{2}{1+z\,\bar{z}}\,\left(\frac{1}{u}\epsilon_{sr}\,C^{\,r}\,\bar{z}^{\frac{2s+1}{2}} +\,\eta_{ij} \left[(j-1) \, J^i - z \, \partial_z J^i \right] \, \bar z^{j+1} \right)\bigg] \,dz^2 \cr
& +\Big(-1+\eta_{ij}\,(-z)^{1-i}\,J^j\Big)\,du^2 - \,\bigg(\,\epsilon_{rs} \, C^{\,r}\,(-z)^{\frac{1-2s}{2}}+ u \,\eta_{ij}\,\partial_z\,((-z)^{1-i}\,J^j)\bigg)\,du\,dz \cr &
\end{align}
\end{itemize}
%
%
Here all the functions $(\kappa, J^i, C^r)$ are holomorphic and the non-zero components of $\eta_{ij}$ are $\eta_{-1 \, +1} =1/2$ and $\eta_{00} =-1$, and $\epsilon_{- \frac{1}{2} \, + \frac{1}{2}} =1$. Now that we have obtained the locally flat solutions with the boundary conditions we wanted as well as imposed a consistent variational problem we can turn to analysing the symmetry algebras of these solution spaces.
%

\section{Asymptotic symmetries}
\label{sec5}
Here we seek all the vector fields that keep us within the classes of solutions (\ref{finalLFr2}, \ref{finalLFs2}) and compute their commutator algebras. 
%
\subsection{The vector fields }
\label{thevectorfields}
From the general analysis of asymptotic symmetries done in (\ref{asv1}-\ref{asv2}) the first few components of the vector fields that keep us within the class of (\ref{finalLFr2}, \ref{finalLFs2}) are given by,
\begin{align}
&\xi^{z}_{(0)}=Y(z),~~\xi^{\bar{z}}_{(0)}=\overline{Y}(z,\bar{z}),~~\xi^{u}(u,z,\bar{z})=\xi^{u}_{(0)}(z,\bar{z})+\frac{u}{2}\,D_a\,V^a \\
&\xi^{r}_{(0)}=-\frac{1}{2}\,D_a\,V^a,~~ \xi^{a}_{(1)}=-g^{ab}_{(0)}\,D_b\,\xi^u \\
&\xi^a_{(2)}=\frac{1}{2}g^{ab}_{(0)}\,g^{(1)}_{bc}\,g^{cd}_{(0)}\,D_d\,\xi^u\\
&\xi^r_{(1)}=\frac{1}{2}\,g^{ab}_{(0)}D_aD_b\,\xi^u-\frac{1}{4}\,\xi^a_{(0)}\,D_a\,\big(g^{bc}_{(0)}g^{(1)}_{bc}\big)-\frac{1}{4}\,\partial_u\,\big(\xi^u\,g^{bc}_{(0)}g^{(1)}_{bc}\big)
\end{align}
In the above solution $V^a=(Y(z),\overline{Y}(z,\bar{z}))$. Under the action of these vector fields the data $g_{ab}^{(0)}$ and $g_{ab}^{(1)}$ transform as follows,
\begin{align}
&\delta g_{ab}^{(0)}=\mathcal{L}_{V^c}g_{ab}^{(0)}-2\,\partial_u\,\xi^u\,g_{ab}^{(0)} \label{bdrymetrictra}\\
&\delta g_{ab}^{(1)}=\mathcal{L}_{V^c}g_{ab}^{(1)}-\partial_u\,\xi^u\,g^{(1)}_{ab}+\xi^u\,\partial_u g_{ab}^{(1)}+g_{ab}^{(0)}\,g^{cd}_{(0)}D_c\,D_d\,\xi^u-2\,D_a\,D_b\,\xi^u \label{cond1}
\end{align}
We now have to impose conditions coming from the variational principle: $g^{(1)}_{\bar{z}\bar{z}}=g^{(1,0)}_{\bar{z}\bar{z}}+u\,g^{(1,1)}_{\bar{z}\bar{z}}=0$ as found in  section \ref{varPolyakov}. Considering $\delta g^{(1)}_{\bar{z}\,\bar{z}}$ component in eq (\ref{cond1}) and setting it to zero gives
\begin{align}
\label{varPimp}
\delta g^{(1,0)}_{\bar{z}\bar{z}}+u\,\delta g^{(1,1)}_{\bar{z}\bar{z}}=-2\,D_{\bar{z}}D_{\bar{z}}\big(\xi^{u}_{(0)}+\frac{u}{2}\,D_a\,V^a\big) =0
\end{align}
This requires setting $D_{\bar{z}}D_{\bar{z}}\,\xi^u_{(0)}=0$ and $D_{\bar{z}}D_{\bar{z}}\,D_a\,V^a=0$ separately.
\begin{itemize}
\item The equation $D_{\bar{z}}D_{\bar{z}}\,\xi^u_{(0)}=0$ can be solved for $\xi^u_{(0)}$ and we obtain
\bea
\label{finalxiu0}
\Sigma_2 &=& {\mathbb R}^2: ~~~
\xi^u_{(0)}= P_{-\frac{1}{2}}(z)+\,\bar{z}\,P_{\frac{1}{2}}(z)  \cr
\Sigma_2 &=& S^2: ~~~ \xi^u_{(0)}=\frac{2}{1+z\bar{z}}\,\left(P_{-\frac{1}{2}}(z)+\,\bar{z}\,P_{\frac{1}{2}}(z) \right) 
\eea
\item The equation  $D_{\bar{z}}D_{\bar{z}}\,D_a\,V^a=0$ allows us to solve for $\overline{Y}(z,\bar{z})$ as
\bea
\label{finalbary}
\overline{Y}(z,\bar{z})=Y_{-1}(z)+\bar{z}\,Y_0(z)+\bar{z}^2\,Y_1(z)
\eea
in either case of $\Sigma_2$. 
\end{itemize}
Therefore, the vector fields that keep us within our solution spaces are characterised by six arbitrary holomorphic functions:  $(Y(z), Y_{-1}(z), Y_0(z), Y_1(z), P_{-1/2}(z), P_{1/2}(z))$. The subleading components of vector fields are then found in terms of these leading order components as mentioned in the general analysis done in Appendix \ref{appendixA}. If these holomorphic functions are allowed to have poles, then the supertranslation vector field $\xi^u_{(0)}$ for a specific choice of $P_{-\frac{1}{2}},P_{\frac{1}{2}} $ as follows, $$P_{-\frac{1}{2}}(z)=-\frac{\left(1+w\bar{w}\right)\,\bar{w}}{2\,(z-w)},~~ P_{\frac{1}{2}}(z)=\frac{\left(1+w\bar{w}\right)}{2\,(z-w)}$$ becomes,
\begin{align}
\xi^u_{(0)}=	\frac{1+w\,\bar{w}}{z-w}\frac{\left(\bar{z}-\bar{w}\right)}{1+z\bar{z}} \label{supertraWI}
\end{align} 
The vector field in equation (\ref{supertraWI}) was used in \cite{Campiglia:2015kxa} to show the equivalence between supertranslation Ward identity and Weinberg leading soft theorem for positive helicity graviton localised at the reference direction $(w,\bar{w})$ on the sphere. The vector field in (\ref{supertraWI}) induces transformation of $ g^{(1;0)}_{\bar{z}\bar{z}}$ which is zero everywhere on the sphere except singular in the direction $(w,\bar{w})$,
\begin{align}
	\delta g^{(1;0)}_{\bar{z}\bar{z}}=-\frac{4\pi\,\left(1+w\,\bar{w}\right)}{(1+z\bar{z})}\delta^{(2)}(z-w). \label{gzz10}
\end{align}  
In a similar spirit for the vector field in equation (\ref{finalbary}) the choice,
\begin{align} Y_{-1}(z)=\frac{\bar{w}^2}{(z-w)}\,,~~Y_{0}=-\frac{2\,\bar{w}}{(z-w)}\,,~~Y_{1}=\frac{1}{(z-w)} \label{Diff}
\end{align}coincides with the vector field used in \cite{campiglia:2015al} to show the equivalence between Diff$(S^2)$ Ward identity and Cazchao Strominger subleading soft theorem for positive helicity graviton (the conjugate vector field give rise to Ward identity that is equivalent to CS soft theorem for negative helicity graviton)\footnote{As mentioned in \cite{Campiglia:2015kxa} this vector field act as a kernel from which any smooth vector field can be constructed and therefore Ward identity associated with any Diff($\mathcal{S}^2$) vector field on the sphere can be obtained from the Ward identity associated with this vector field}. For this particular choice of vector field,
\begin{align}
	\delta g^{(1;1)}_{\bar{z}\bar{z}}=4\pi \delta^{(2)}(z-w) \label{gzz11}
\end{align}
The singularity in (\ref{gzz10}) and (\ref{gzz11}) is similar in nature to the one generated by superrotation vector fields of the type 
\begin{align}
	Y^z=\frac{1}{z-w},\label{superrotation}
\end{align} which change the sphere metric according to (\ref{bdrymetrictra}) at null infinity by adding singularities at isolated points \cite{Strominger:2016wns}.
%
%
\subsection{Algebra of vector fields}
Now that we have our full set of vector fields that enable us to move in the space of solutions (\ref{finalLFr2}, \ref{finalLFs2}), one can check that the Lie brackets between any two of these vector fields close under the modified commutator (see for instance \cite{barnich:2010tr})
\bea
[\,\xi_1\,,\,\xi_2\,]_M :=\,\xi_1^{\sigma}\,\partial_{\sigma}\,\xi_2-\,\xi_2^{\sigma}\,\partial_{\sigma}\,\xi_1-\,\delta_{\xi_1}\,\xi_2+\,\delta_{\xi_2}\,\xi_1 \label{lie brac}
\eea
Calculating the bracket as in eq (\ref{lie brac}) we find:
\begin{align}
&[\,\xi_1\,,\,\xi_2\,]_M=\bigg(\hat{\xi}^u_{(0)}+\frac{u}{2}\,D_a\,\hat{V}^a\bigg)\,\partial_u+\bigg(\hat{V}^a-\frac{1}{2\,r}\,g_0^{ab}\,D_b\,\hat{\xi}^u\bigg)\,\partial_a-\frac{r}{2}\,D_a\,\hat{V}^a\,\partial_r+ \cdots
\end{align}
where
\bea
\hat{\xi}_{(0)}^u = V_1^a\,\partial_a\,\xi_{2\,(0)}^{u}+\frac{1}{2}\,\xi_{1\,(0)}^{u}\,D_a\,V^a_2-V_2^a\,\partial_a\,\xi_{1\,(0)}^{u}-\frac{1}{2}\,\xi_{2\,(0)}^{u}\,D_a\,V^a_1\nonumber
\eea
\bea
\hat{V}^a = V^b_1\,\partial_b\,V^a_2-V^b_2\,\partial_b\,V^a_1
\eea
Considering the Lie bracket with $V_1^a=V_2^a=0$ one can check that the resultant $\hat{\xi}^u_{(0)}=0$ -- thus our supertranslation vector fields commute with each other. Any of the asymptotic vector fields is determined by $(Y(z), \overline{Y} (z, \bar z), \xi^u_{(0)}(z, \bar z))$ with $\overline Y$ and $\xi_{(0)}^u$ given in (\ref{finalxiu0}, \ref{finalbary}). 
\begin{itemize}
\item We denote the vector fields with $Y(z) = \xi^u_{(0)}(z, \bar z) =0$ and $\overline{Y} (z, \bar z) = - z^n \bar z^{i+1}$ by ${\cal J}^i_n$ for $i \in \{ -1, 0, 1\}$ and $n \in {\mathbb Z}$. 
%
The commutator of these vector fields is given by,
\begin{align}
\label{sl2alg1}
[{\cal J}^{i}_m\,,\, { \cal J}^{j}_{n}\,]=\big(i-j \big)\,{\cal J}^{i+j}_{m+n}\
\end{align}
This can be recognised as the $\mathfrak{sl}_2$ current algebra. 
\item Denoting the vector field with $\overline Y = \xi^u_{(0)} =0$ and $Y(z) = -z^{n+1}$ by ${\cal L}_n$ for $n\in {\mathbb Z}$ we find: 
\begin{align}
\label{sl2alg2}
[{\cal L}_m, {\cal L}_n] = (m-n) \, {\cal L}_{m+n}, ~~~ [\,{\cal L}_m\,,{\cal J}^{a}_n\,]=-n\,{\cal J}^a_{m+n}
\end{align}
The first of these is simply the Witt algebra. 
\item Finally denoting the vector field with $Y = \overline{Y} =0$ and  
\bea
\xi^u_{(0)} = - z^{r+\frac{1}{2}} \, \bar{z}^{s+\frac{1}{2}}  ~~ {\rm for}~~ \Sigma_2 = {\mathbb R}^2, ~~~
\xi^u_{(0)} = -\frac{z^{r+\frac{1}{2}}\bar{z}^{s+\frac{1}{2}}}{1+z\bar{z}}  ~~{\rm for} ~~ \Sigma_2 = S^2
\eea
by ${\cal P}_{r,s}$ where $r \in {\mathbb Z}+\frac{1}{2}$ and $s\in \{ - \frac{1}{2}, \frac{1}{2}\}$, the remaining commutators work out to be:
\bea
\label{sl2alg3}
[{\cal L}_n, {\cal P}_{r,s}] = \frac{1}{2} (n-2 \, r) \, {\cal P}_{n+r, s}, ~~ [{\cal J}^a_n, {\cal P}_{r,s}] = \frac{1}{2} (a-2 \, s) \, {\cal P}_{n+r, a+s}, ~~ [{\cal P}_{r,s}, {\cal P}_{r',s'}] =0 
\eea
\end{itemize}
Thus the final result for the algebra of vector fields that preserve our spaces of locally flat solutions (\ref{finalLFr2}, \ref{finalLFs2}) are (\ref{sl2alg1}, \ref{sl2alg2}, \ref{sl2alg3}). This algebra is identical to the one uncovered from an analysis of conformal soft theorem in the celestial CFT by \cite{banerjee:2020gp, Banerjee:2021dlm}. 
%
\subsection{Variations of fields in locally flat solutions}
Our final set of locally flat solution (both for $\Sigma_2 = {\mathbb R}^2$ or $S^2$) are characterised by six holomorphic fields: $(\kappa(z), J^i(z), C_r(z))$. For the sake of completeness here we compute how these transform under the generators of the symmetry algebra found in the previous subsection. It turns out that the transformations of the fields are form-invariant for either of the solutions (\ref{finalLFr2}) and (\ref{finalLFs2}). So we will demonstrate this for the case of $\Sigma_2 = {\mathbb R}^2$. First we rewrite the field  $C^r$ as (like in \cite{campiglia:2020jp, Campiglia:2021bap} for instance) 
\begin{align}C^r(z)=-2\,\partial_z ^2\,\mathcal{C}^r(z)+\bigg(\frac{1}{2}\,\eta_{ij}\,J^i(z)\,J^j(z)-\kappa(z)\bigg)\,\mathcal{C}^r(z)+g^r_{i\,s}\,\big(-\partial\,J^i(z)\,\mathcal{C}^s(z)-2\,J^i(z)\,\partial \mathcal{C}^s(z)\big). \label{cr}
\end{align}
Here the non-zero structure constant $g^r_{is}$ are  given by: $g^{\frac{1}{2}}_{0\,\frac{1}{2}}=g^{-\frac{1}{2}}_{-1\,\frac{1}{2}}=1, g^{\frac{1}{2}}_{1\,-\frac{1}{2}}=g^{-\frac{1}{2}}_{0\,-\frac{1}{2}}=-1$. 
We can re-write the vector field that preserves the form of our solutions (up to $1/r$-order terms):
\bea
&\xi^u(u,z,\bar{z})=\epsilon_{rs}\,\bar{z}^{\frac{2r+1}{2}}\,P^s(z)+\frac{u}{2}\,\big(\eta_{ij}\,(1+j)\,Y^i(z)\,\bar{z}^j+\partial_z  Y(z)\big)\\
\nonumber \\
&\xi^r(r,u,z,\bar{z}) = -\frac{r}{2}\,\big(\eta_{ij}\,(1+j)\,Y^i(z)\,\bar{z}^j+\partial_z  Y(z)\big)+\eta_{ij}\,(1+j)\,J^i(z)\,\bar{z}^j\,\big(-\epsilon_{rs}\,\,(2s+1)\,P^r(z)\,\bar{z}^{\frac{2s-1}{2}} \nonumber \\
\nonumber \\
& +u\,\eta_{ij}\,(1+j)\,j\,Y^i(z)\,\bar{z}^{j-1}\big) - \epsilon_{rs}\,(2s+1)\,\partial_z\,P^r(z)\,\bar{z}^{\frac{2s-1}{2}}+u\,\eta_{ij}\,(1+j)\,j\,\partial_z Y^i(z)\,\bar{z}^{j-1}\nonumber \\
\nonumber \\
&-\frac{1}{r}\,\left[\epsilon_{rs}\,\tfrac{2s+1}{2}\,\bar{z}^{\frac{2s-1}{2}}\,(-2\,P^r(z)+C^r(z))+u\,\eta_{ij}\,(1+j)\,j\,\bar{z}^{j-1}\,(Y^i(z)+\partial_z J^i(z))\right] + \cdots
\eea
\begin{align}
\xi^z(r,u,z,\bar{z})&=Y(z)+\frac{2}{r}\,\left[ \epsilon_{rs}\,\frac{2s+1}{2}\,P^r(z)\,\bar{z}^{\frac{2s-1}{2}}-u\,\eta_{ij}\,(1+j)\,j\,Y^i(z)\,\bar{z}^{j-1}\right] + \cdots \cr
\xi^{\bar{z}}(r,u,z,\bar{z})&=\eta_{ij}\,Y^i(z)\,\bar{z}^{1+j}+\frac{1}{r}\,\Big[2\,\eta_{ij}\,J^i(z)\,\bar{z}^{1+j}\,\epsilon_{rs}\,(2s+1)\,P^r(z)\,\bar{z}^{\frac{2s-1}{2}}+2\,\epsilon_{rs}\,\partial_z\,P^r(z)\,\bar{z}^{\frac{2s+1}{2}} \nonumber \\&-u\,(2\,\eta_{ij}\,J^i(z)\,\bar{z}^{1+j}\,\eta_{kl}\,(1+l)\,l\, Y^k(z)\,\bar{z}^{l-1}+\eta_{ij}\,(1+j)\,\partial_z Y^i(z)\,\bar{z}^j+\partial_z^2\,Y(z)\,\big)\Big] + \cdots
\end{align}
Under these residual diffeomorphism the fields in the solution space are transformed as follows,
\bea
\label{deltaphis}
\delta J^i (z)&=& J^{i}(z)\,\partial\,Y(z)+Y(z)\,\partial\,J^i(z)+\, {f^{i}}_{jk}\,J^j(z)\,Y^k(z)-\partial\,Y^i(z) \cr
\delta \kappa(z) &=& Y(z)\,\partial\,\kappa(z)+2\,\kappa(z)\,\partial Y(z)+\partial^3\,Y(z) \cr
\delta \mathcal{C}^r (z)&=& Y(z)\,\partial\,\mathcal{C}^r(z)-\frac{1}{2}\,\partial Y(z)\,\mathcal{C}^r(z)+{g^r}_{i s}\,\frac{1}{2}\,Y^i(z)\, \mathcal{C}^s(z)+P^r(z)
\eea
The values of the non-zero components of the structure constants ${f^i}_{jk} = -{f^i}_{kj}$ are $ f^{-1}_{\,-1\,\,0}=-1\,,~f^{0}_{\,\,1\,,-1}=\frac{1}{2}\,,~f^{1}_{\,\,1\,,0}=1$. Even though we have presented the results of the variations of the fields in the case of $\Sigma_2 = {\mathbb R}^2$ the computations can be repeated in the case of $\Sigma_2 = S^2$ and we find that the expression for the variations take the same form as in (\ref{deltaphis}). So we will not provide the details of that exercise here.

One expects that these $\delta_\xi \{ \kappa(z), J^i(z), {\cal C}^r(z)\}$ are given by the Poisson brackets of the charges $Q[\xi]$ corresponding to the vector field $\xi$ with the fields. Assuming this interpretation to be true, one can comment on the nature of these fields by looking at their transformations (\ref{deltaphis}). The $J^i(z)$ transforms as a primary of weight $ (h=1)$ under left Virasaro transformation generator $Q[Y(z)]$. The field $\kappa(z)$ transforms like a quasi primary of weight $(h=2)$ with an inhomogeneous term that is similar to the chiral stress tensor component of a $2d$ CFT. The $\mathcal{C}^r(z)$ are related to supertranslation current and transform like the primary of weight $(h=-\frac{1}{2})$. The  fields $\kappa(z), J^i(z), {\cal C}^r(z)$ also transform as Kac-Moody primaries with $j=0,1, \frac{1}{2}$ respectively under the $\widehat{\mathfrak{sl}_2}$ algebra.

As we saw in section \ref{pg-lf-solns} after imposing variational problem $g^{(1)}_{\bar{z}\bar{z}}=0$ the locally flat solutions are given in terms of  fields, $$\{g_1(z),g_2(z),g_3(z),g_4(z),\upsilon_{(1)}(z),\upsilon_{(2)}(z),\zeta(z)\}.$$  with the condition $g_1(z)\,g_4(z)-g_3(z)g_2(z)=1$. This condition may be used to eliminate $g_1(z)$ (whenever it is non-zero) in terms of other three fields. These fields are more fundamental as $(\kappa(z), J^i(z), C_r(z))$ are defined in term of these fields (see \ref{fzz}, \ref{crs}). We can find the transformations of these more fundamental fields under the action of the vector fields specified by $(Y(z), Y_i(z), P_r(z))$ and we obtain:
%
%
\begin{align}
&\delta g_2(z)=Y(z)\,\partial_z\,g_2(z)+g_1(z)Y_{-1}(z)\,-\frac{1}{2}\,g_2(z)\,Y_0(z)\cr
&\delta g_3(z)=Y(z)\,\partial_z\,g_3(z)-g_4(z)Y_{1}(z)\,+\frac{1}{2}\,g_3(z)\,Y_0(z) \cr
&\delta g_4(z)=Y(z)\,\partial_z\,g_4(z)+g_3(z)Y_{-1}(z)\,-\frac{1}{2}\,g_4(z)\,Y_0(z) \cr
&\delta \zeta(z)= Y(z)\,\partial_z\,\zeta (z)
\end{align}
%
%
Therefore, the fields $g_2(z),g_3(z),g_4(z)$ transform like scalars under left Virasaro transformation $Y(z)$. To write the transformation properties of $\upsilon_{(1)}$ and $\upsilon_{(2)}$ we define:
\begin{align}
\tilde{\upsilon}_{(1)}(z)=g_4(z)\,\frac{\upsilon_{(1)}}{\sqrt{\partial_z\,\zeta (z)}}+g_2(z)\,\frac{\upsilon_{(2)}}{\sqrt{\partial_z\,\zeta(z)}}\cr
\tilde{\upsilon}_{(2)}(z)=g_3(z)\,\frac{\upsilon_{(1)}}{\sqrt{\partial_z\,\zeta(z)}}+g_1(z)\,\frac{\upsilon_{(2)}}{\sqrt{\partial_z\,\zeta (z)}}
\end{align}
Then we find that $\tilde{\upsilon}_{(1)}(z),\tilde{\upsilon}_{(2)}(z)$ transform as:
\bea
\delta \tilde{\upsilon}_{(1)}(z) &=& Y(z)\,\partial_z \tilde{\upsilon}_{(1)}(z)-\frac{1}{2}\,\partial_z\,Y(z)\,\tilde{\upsilon}_{(1)}(z)+Y_{-1}(z)\,\tilde{\upsilon}_{(2)}(z)-\frac{1}{2}\,Y_{0}(z)\,\tilde{\upsilon}_{(1)}(z)+P_{-\frac{1}{2}}(z)\cr
\delta \tilde{\upsilon}_{(2)}(z)&=&Y(z)\,\partial_z \tilde{\upsilon}_{(2)}(z)-\frac{1}{2}\,\partial_z\,Y(z)\,\tilde{\upsilon}_{(2)}(z)-Y_{1}(z)\,\tilde{\upsilon}_{(1)}(z)+\frac{1}{2}\,Y_{0}(z)\,\tilde{\upsilon}_{(2)}(z)+P_{\frac{1}{2}}(z) \cr &&
\eea
\subsection{Comments on the charges}

 The locally flat solutions that we have obtained parametrise the space of gravitational vacua. The fields $\{\kappa(z), J^i(z),\mathcal{C}^r(z)\}$ can be thought of as  `Goldstone' modes associated with  spontaneous symmetry breaking of  Virasaro, chiral $\mathfrak{sl}_2$ current and $\mathfrak{u(1)}$ current algebra symmetry. In order to construct complete phase space we are required to construct soft charges   that generate symmetry transformation on these fields. From what we have seen the fields in the solutions actually behave like (quasi-) primaries under the chiral Virasoro generators as well as the current algebra ones and thus one expect the charges to be given more appropriately as line integrals as it is the norm for a chiral conformal field theory. It is not clear how to obtain such line integral charges from the 4$d$ bulk perspective where the usual route of defining charges through covariant phase space formalism \cite{Wald} or Barnich-Brandt method \cite{Barnich} lead to co-dimension two surface charges that are defined on the celestial sphere $\mathcal{S}^2$.
 
In \cite{compere:2018fr} charges for gravitational vacua exhibiting supertranslation and Diff $(\mathcal{S}^2)$ symmetry were constructed  and it was shown that these co-dimension two charges corresponding to supertranslations  vanish but charges for superrotations which are proportional to terms quadratic in $g^{(1,0)}_{ab}$ and their derivatives are in general non-vanishing and conserved. These charges evaluate to zero for our chiral solutions in (\ref{finalLFr2}, \ref{finalLFs2}).  Therefore one needs to find an alternative prescription  to calculate charges that are line integrals and provide representation of chiral $\widehat{\mathfrak{sl}}_2$ algebra and generate correct symmetry transformation on solution space.

 In an ongoing work which we will report in the near future we find that to get the charges as line integrals one can switch to the first order formalism of gravity where these locally flat solutions can be written as connections  $A_{\mu}$ that takes value in $\mathfrak{iso}(1,3)$ algebra. The condition of vanishing curvature for this connection  $(F_{\mu \nu}=0)$ is equivalent to local flatness condition that we used to solve in the 4$d$ Einstein gravity. One can gauge away the $r$ dependence using gauge transformation and get an effective 3$d$ flat connection. The sector within this 3$d$ gauge connection  parametrised by fields $\{\kappa(z), J^i(z)\} $  find a natural interpretation independent of field $\mathcal{C}^r(z)$ in terms of 3$d$ flat $\mathfrak{sl(2,\mathbb{C})}$ Chern-Simons connection. 
 The residual gauge transformations that preserve the form of this connection are a subset of the asymptotic symmetries found in previous section (the $\widehat{\mathfrak{sl}}_2$ algebra and the Witt algebra).\footnote{A similar computation was done in \cite{Nguyen:2020hot} where the Goldstone mode associated with Virasaro transformation and Diff$(S^2)$ superrotation is described by two dimensional  Alekseev-Shatashvili theory obtained after Hamiltonian reduction of 3$d$ Chern-Simons theory that parameterises the gravitational vacua associated with these symmetry transformations.}
 
The charges generating residual gauge transformations in a Chern-Simons theory are line integral given by,
\begin{equation}
\cancel{\delta} Q= \frac{1}{2\pi i}\int dx^{\mu}\,{\rm Tr} (\Lambda\, \delta A_{\mu}) \label{lineint1} ,    
\end{equation} 
where $\Lambda$ is the gauge parameter and $\delta A= d\Lambda +[\Lambda,A]$. Using the formula (\ref{lineint1}) the charges generating chiral $\widehat{\mathfrak{sl}}_2$ current and Virasaro symmetry are given by
$$Q=\frac{1}{2\pi i}\oint dz \, [\eta_{ij}\,Y^i(z)\,J^j(z)+ Y(z)\kappa(z)].$$

The part of gauge connection that has the information about the mode $\mathcal{C}^r(z)$ associated with supertranslation is gauged by the momentum generators $(P_i)$ of Poincare algebra and if one attempts to calculate  charges associated with supertranslation using (\ref{lineint1}) one has to consider full $\mathfrak{iso}(1,3)$ gauge connection as $\delta \! A_{\mu}$  gets non-trivial contribution from the commutators of $\mathfrak{sl(2,\mathbb{C})}$ generators with $P_i$. Since Poincare algebra does not have a non-degenerate bilinear invariant, therefore computation of charges for  $\mathfrak{u(1)}$ currents is not possible using the charge formula (\ref{lineint1}).

Another possible way to obtain these supertranslation charges could be to start with the $\mathfrak{so}(2,3)$ algebra which does have a non-degenerate bilinear invariant. The analogous flat connections in this case describe locally $AdS_4$ solutions having the chiral fields $\{\kappa(z),J^{i}(z),\mathcal{C}^r(z)\}$. These solutions in the flat space limit, {\it i.e.,} when the radius of $AdS_4$ goes to infinity, match with the corresponding locally flat solutions. After obtaining charges as line integrals one can write down a set of holomorphic operator product expansions between the charge generators and the fields. We hope to get some insight into the supertranslation charges in the ${\mathbb R}^{1,3}$ context via a suitable Inonu-Wigner contraction of this computation. This will also allow us to find the central extensions (if any) of the algebra of charges that cannot be seen from the commutator algebra of the corresponding vector fields. This work is currently in progress.
\section{Conclusion and Discussions}
\label{sec6}
In this paper we introduced a new set of boundary conditions for the four dimensional classical gravity without cosmological constant (the ${\mathbb R}^{1,3}$ or ${\mathbb R}^{2,2}$ gravity) that are asymptotically locally flat near the null infinities. Our boundary conditions are non-Dirichlet.  By considering a Chiral gauge for the metric on 2-dimensional spatial manifold $\Sigma_2$ motivated by Polyakov's gauge for asymptotically locally $AdS_3$ of \cite{Avery:2013dja} we provide the complete class of locally flat solutions. Constructing and making use of a set of appropriate boundary terms to impose a variational principle, we restrict the class of locally flat solutions to a subset. By studying the symmetry algebra of the vector fields that leave our final solution space invariant, we show that it is a novel extension of the Poincare algebra that includes a chiral $\mathfrak{sl}_2$ current algebra. Our symmetry algebra is identical to the recently uncovered symmetry algebra of the celestial CFT from its conformal soft theorems \cite{banerjee:2020gp, Banerjee:2021dlm}. 

Our variational problem is defined for $4d$ Einstein gravity with a specific set of boundary terms we proposed in the text. To ensure $\delta S =0$ for the allowed classical solutions we have chosen to impose additional boundary conditions on the configurations that already solve the bulk equations of motion.\footnote{It is known that a consistent variational problem at null-infinities would not allow solutions representing gravitational radiation at infinity. Our boundary conditions, however, are weak enough to at least allow for Schwarzchild black hole in the solution space.}  We would like to emphasise that even in the case of Dirichlet boundary conditions (in which the metric on $\Sigma_2$ is held fixed) there are additional conditions imposed by our variational principle and solving them in a non-chiral fashion requires us to set $g^{(1,1)}_{zz} = g^{(1,1)}_{\bar z \bar z} =0$. Even though we did not provide the details it can be seen easily that the residual large diffeomorphisms of such Dirichlet class of solutions does not permit extension of the Lorentz algebra part of the $\mathfrak{bms}_4$ into two copies of Witt algebra. Thus our boundary conditions would not allow the extended $\mathfrak{bms}_4$ as the asymptotic symmetry algebra in the Dirichlet case. It will be important to explore the consequences of this fact further. 

We provided exact solutions that are locally flat spacetimes. Working in the NU gauge these are polynomials in radial coordinate $r$. When the boundary metric is taken in the Chiral Polyakov gauge, the solution spaces are necessarily complex. Starting from our solution space we can generate three more classes of geometries by $T: u \leftrightarrow -v$ and/or $P: z \leftrightarrow -1/\bar z$ -- which are time-reversal and spatial parity transformations respectively in the asymptotic flat spacetime (with $\Sigma_2 = S^2$). We would like to think of these four classes of solutions as valid in four different coordinate patches of the fully extended (locally flat) geometries. Then these four patches are related to each other by the (P, T) transformations. One can in principle patch together these solutions that are (C) PT invariant. Just as there are four classes of solutions there are four corresponding chiral algebras. This is similar to the two copies of $\mathfrak{bms}_4$ symmetry algebras in the context of Dirichlet boundary conditions -- one at the past null infinity and the other at the future null infinity. Just as an appropriate combination was found by Strominger (see \cite{Strominger:2017zoo} for a review) picked by the CPT invariance of the scattering amplitudes we expect that an appropriate combination of the four copies of our current algebras as well to emerge as the correct symmetry of the scattering amplitudes. 

The locally flat solutions (\ref{finalLFr2}) and (\ref{finalLFs2}) that we arrived at (and worked with) after imposing chiral boundary conditions are indeed complex in the $\mathbb{R}^{1,3}$ gravity (they are real however in the $\mathbb{R}^{2,2}$ gravity as $(z, \bar z)$ will be light-cone coordinates $(x^+, x^-)$)\footnote{Gravity scattering amplitudes in such signature were explored recently by \cite{Atanasov:2021oyu}.}. As a consequence the vector fields generating the asymptotic symmetries of this class of geometries are also complex. Since for LF solutions the Riemann tensor vanishes, in principle these solutions can all be obtained (in open patches) by finite complex coordinate transformation of metric $\eta_{\mu\nu}$ of the Minkowski spacetime $\mathbb{R}^{1,3}$. Expanding the LF solutions around $\eta_{\mu\nu}$ to the first order in the six holomorphic functions the perturbation $h_{\mu\nu}$ can be written as pure diffeomorphisms with the same complex vector fields that generate our asymptotic symmetries. In \cite{Donnay:2020guq} it was shown that  the conformal basis for graviton wave-functions $h^{\Delta, \pm }_{\mu\nu;a}$ of definite helicity (where $\pm$ superscript denotes whether the graviton is incoming/out-going and the subscript $a$ denotes its helicity) in the soft limits $\Delta \rightarrow 1$,  $\Delta \rightarrow 0$ become pure diffeomorphisms: $h^{\Delta, \pm }_{\mu\nu;a} = \nabla_\mu \zeta^{\Delta, \pm }_{\nu;a} + \nabla_\nu \zeta^{\Delta, \pm }_{\mu;a}$ \footnote{Even though the analysis in \cite{Donnay:2020guq} was done in harmonic gauge, the boundary vector fields that generate generalised BMS transformations at leading order is same as that of Bondi gauge and hence NU gauge. }. Near null infinity these vector fields $\zeta^{\Delta}_{\mu;\pm}$ in the soft limit become generator of supertranslation symmetry and Diff$(\mathcal{S}^2)$ symmetry and for positive helicity soft graviton take the form of (\ref{supertraWI}), (\ref{Diff}) respectively .  The conformal primary wave functions given in terms of these vector fields are then interpreted as  Goldstone modes of spontaneously broken asymptotic symmetries of gravity theory\footnote{ In fact our solution spaces contain one additional holomorphic function $\kappa(z)$ associated with the vector field in (\ref{superrotation}) which is possibly the Goldstone mode of addition positive soft helicity mode (related to the shadow transform of negative helicity graviton) with $\tilde{\Delta}=2$ in section 5.3.1 of \cite{Donnay:2020guq} }. Similarly one can interpret our complex solutions as the `condensates' of soft gravitons of definite helicity-positive (negative) if the solutions are characterised by six holomorphic(anti-holomorphic) functions.

  In this paper,  we use the definition of variational principle in the conventional sense that implies stationarity of the action $S_{EH+bdry.}$ on the solutions. As argued in \cite{Fiorucci:2021pha}, one modifies the definition of a well-defined variational principle in the presence of radiation to allow for some presymplectic
flux through the boundary. However, we believe that even if we consider radiative solutions, then in the asymptotic regions unaffected by the radiation (for instance, if a pulse of radiation reaches the boundary during some finite interval in $u$, then sufficiently far away from this interval), the solutions should approach one of the vacuum solutions. And these vacuum solutions, we posit, have to be one of the classes of locally flat solutions for which the total action is stationary. It will be interesting to see whether our
symmetry analysis still holds in the case of presymplectic flux,
through the boundary, instead of the stationarity condition of the action.
  
A bigger symmetry algebra $(w_{1+\infty})$ than the one we uncovered in this paper is observed from the conformal soft limits of graviton operators in the celestial CFT \cite{guevara:2021hps, Strominger:2021lvk} recently. It will be interesting to explore if our boundary conditions can be generalised to incorporate these extended symmetries or not.\footnote{See \cite{Freidel:2021ytz} for the derivation of charges using vacuum Einstein equations that form the canonical representation of these $w_{1+\infty}$ algebra.} Following the analysis of \cite{Campiglia:2016efb, Donnay:2022sdg}, one expects that the vector fields generating $w_{1+\infty}$ are overleading in $r$ as one approaches the boundary of spacetime. However, it can be checked from eq. (\ref{vecasymp}) that such behaviour for vector fields in the NU gauge does not exist. Therefore, to extend the analysis of algebra in our present work to $w_{1+\infty}$ algebra, one has to solve for vector fields in a different gauge. In \cite{Donnay:2020guq}, the authors derived the vector fields that generate arbitrary diffeomorphisms of the celestial sphere in harmonic gauge. One can do a similar analysis as theirs by demanding less restrictive boundary conditions that allow for metric perturbation around Minkowski spacetime to fall off as $\mathcal{O}(r^n)$ for $n\geq 2$ and obtain vector fields that are overleading in $r$ near null infinity. It is plausible that imposing a variational problem for such configuration  might further restrict the solution space, the residual gauge transformations of which could map to $w_{1+\infty}$. However, the interpretation of solutions with such overleading behaviour in $r$ near the null infinity from the spacetime perspective is unclear as they do not obey the standard asymptotically flat boundary conditions.

We considered the metric on $\Sigma_2$ in the Polyakov gauge. The other natural choice is the conformal gauge in which we have found a class of locally flat solutions (see Appendix \ref{appendixB} for details). We are currently analysing the symmetry algebra of these solution spaces and what these would mean for the scattering amplitudes of gravitons. 

%


The Chiral gravity boundary conditions of \cite{Avery:2013dja} that gave rise to the Polyakov's chiral $\mathfrak{sl}_2$ algebra from the $AdS_3$ gravity was generalised to the supersymmetric context in \cite{Poojary:2017xgn}. It should be possible to generalise the boundary conditions of this paper to $4d$ gravitational theories with supersymmetry to find supersymmetric extensions of the Chiral $\mathfrak{sl}_2$ current algebra found here. Very recently such supersymmetric extensions have been observed in the celestial amplitudes \cite{Jiang:2021ovh}. 

In \cite{gupta:2020ns} 3-dimensional scalar field theories coupled to Carrollian geometries were constructed that were manifestly invariant under a set of Carroll diffeomorphisms and Weyl transformations. Gauge fixing those symmetries lead to field theories with $\mathfrak{bms}_4$ symmetries. It will be interesting to see if a gauge fixing similar to the one considered here leads to the emergence of $\mathfrak{sl}_2$ current algebra in such theories or not. 

We have obtained our chiral algebra from the future null-infinity. The generalised $\mathfrak{bms}_4$ has been shown to emerge from the asymptotic symmetry analysis around the time-like infinity as well \cite{AH:2020rfq}. It will be interesting to analyse our problem from the point of view of the time-like infinity. 


\section*{Acknowledgements}

We would like to thank Aneesh PB and Alok Laddha for helpful conversations about and interest in this work. The work of PP is partially supported by a grant to CMI from the Infosys Foundation.
 
\appendix
\section{The residual symmetries in NU gauge}
\label{appendixA}

Here we consider vector fields $\xi^\mu \partial_\mu$ which leave our boundary conditions and gauge choice invariant. First imposing the gauge conditions $\delta g_{ru} = \delta g_{rr} = \delta g_{ra} =0$ require that the vector field components satisfy:
\bea
\label{asv1}
\partial_r\xi^u = 0, ~~ \partial_u \xi^u + \partial_r \xi^r = g_{ua} \, \partial_r \xi^a, ~~\partial_a \xi^u = g_{ab} \partial_r \xi^b
\eea
These equations can again be solved in an asymptotic expansion around large-$r$. Expanding $\xi^r$ and $\xi^a$ as
\bea
 \xi^r = \sum_{n=0}^\infty r^{1-n} \xi^r_{(n)} (u,z,\bar z), ~~ \xi^a = \sum_{n=0}^\infty r^{-n} \xi^a_{(n)} (u,z,\bar z) \label{vecasymp}
\eea
and imposing the conditions obtained from form-invariance of our class of geometries ($g^{(0)}_{ua} = g^{(0)}_{uu} = g^{(1)}_{ua} = g^{(1)}_{uu} =0$) leads to the following conditions on the first few coefficients:
\bea
\label{asv2}
\xi^u (u,z, \bar z) &=& \xi^u_{(0)}(z, \bar z) + u \, \xi^u_{(1)} (z, \bar z), ~~ \xi^a_{(0)} = \xi^a_{(0)}(z, \bar z) \cr
\xi^r_{(0)} &=& - \xi^u_{(1)}, ~~ \xi^a_{(1)} = - g_{(0)}^{ab} D_b \xi^u, ~~ \xi^a_{(2)} = \frac{1}{2} g_{(0)}^{ab} g^{(1)}_{bc} g_{(0)}^{cd} D_d \xi^u
\eea
and so on. At the leading order there are four undetermined functions in these vector fields: $\{\xi^u_{(0)}(z, \bar z), \xi^u_{(1)} (z, \bar z), \xi^a_{(0)} (z, \bar z) \}$. If we further assume that the variation of $g_{(0)}^{ab} g^{(1)}_{ab}$ vanishes leads to determination of $\xi^r_{(1)}$ as well to be:
\bea
\label{asv3}
\xi^r_{(1)} = \frac{1}{2} \square \, \xi^u - \frac{1}{4} \xi^a_{(0)} D_a \left(g_{(0)}^{bc} g^{(1)}_{bc} \right) - \frac{1}{4} \partial_u \left( \xi^u \, g_{(0)}^{bc} g^{(1)}_{bc}  \right)
\eea
As in [Barnich] we assume the gauge choice $g_{(0)}^{ab} g^{(1)}_{ab}=0$ for the background geometries. This choice of $\xi^r_{(1)}$ also ensures that the transformed $g^{(1)}$ is also traceless and remains at most linear in $u$. Under such a diffeomorphism the data $g^{(0)}_{ab}$, $g^{(1)}_{ab}$ etc transform as follows:
\bea
\delta g^{(0)}_{ab} = {\cal L}_{\xi_{(0)}^a} g^{(0)}_{ab} - 2 \, \partial_u \xi^u \, g^{(0)}_{ab}
\eea
which means that part of the bulk diffeomorphism acts as a diffeo and a Weyl transformation of $2d$ geometry $g^{(0)}_{ab}$. Then the data at ${\cal O}(r)$ transforms as:
\bea
\delta g^{(1)}_{ab} &=& - \partial_u \xi^u g^{(1)}_{ab} + \xi^u \, \partial_u g^{(1)}_{ab} + \xi^c D_c g^{(1)}_{ab} + g^{(1)}_{ac} D_b \xi^c + g^{(1)}_{bc} D_a \xi^c \cr
&& + g^{(0)}_{ab} \, \square \xi^u - 2 \, D_a D_b \xi^u 
\eea
where the r.h.s, as can be observed, is also traceless when $g^{(1)}$ is traceless, and at most linear in $u$ as claimed. 

\section{Locally flat solutions in Conformal gauge}
\label{appendixB}
This gauge corresponds to setting $f= \bar f =0$ and the locally flat solutions found in the text reduce to
\bea
\label{lf-cg}
g^{(0)}_{zz} &=& g^{(0)}_{\bar z \bar z} =0, ~~ g^{(0)}_{z \bar z} = \frac{1}{2} \zeta(z, \bar z), ~~ g^{(1)}_{z \bar z} =0, ~~  \cr
g^{(1,1)}_{zz} &=& \kappa(z) + \frac{\partial_z^2 \zeta(z, \bar z)}{\zeta(z, \bar z)} - \frac{3 \, (\partial_z\zeta(z, \bar z))^2}{2 \, (\zeta(z, \bar z))^2}, \cr
g^{(1,1)}_{\bar z \bar z} &=& \bar \kappa (\bar z) + \frac{\partial_{\bar z}^2 \zeta(z, \bar z)}{\zeta(z, \bar z)} - \frac{3 \, (\partial_{\bar z}\zeta(z, \bar z))^2}{2 \, (\zeta(z, \bar z))^2} \cr
g^{(1,0)}_{z z} &=& -2  \left( \partial_z^2 C(z, \bar z) - \frac{ \partial_z C(z, \bar z) \, \partial_z \zeta(z, \bar z)}{\zeta(z, \bar z)} \right)- g^{(1,1)}_{zz} \, C(z, \bar z), \cr
g^{(1,0)}_{\bar z \bar z} &=& -2  \left( \partial_{\bar z}^2 C(z, \bar z) - \frac{ \partial_{\bar z} C(z, \bar z) \, \partial_{\bar z} \zeta(z, \bar z)}{\zeta(z, \bar z)} \right)- g^{(1,1)}_{\bar z \bar z} \, C(z, \bar z) \cr &&
\eea
where $C$ and $\zeta$ are arbitrary functions of both $z$ and $\bar z$. Thus an element of this space of solutions is specified by the functions
\bea
(\kappa(z), ~ \bar \kappa(\bar z), ~ C(z, \bar z), ~ \zeta(z, \bar z))
\eea
The $\zeta$ dependent terms in $g^{(1,1)}_{zz}$ and $g^{(1,1)}_{\bar z \bar z}$ are simply the Schwarzian derivatives of the conformal factor $\zeta$ and they vanish if and only if $\zeta (z, \bar z) = \frac{\alpha^2}{(\beta + z \bar z)^2}$. The cases of the boundary metric being ${\mathbb R}^2$ corresponds to $\alpha = \beta \rightarrow \infty$ (that is, $\zeta =1$) whereas the case of round unit $S^2$ corresponds to $\alpha =2$ and $\beta =1$, and $\alpha =2$ and $\beta =-1$ corresponds to the case of ${\mathbb H}^2$. This space of solutions gets constrained further by imposition of a variational principle. 
%
\subsection{Imposing variational principle}
In this case we have $\delta g^{(0)}_{ab} = g^{(0)}_{ab} \, \zeta^{-1} \, \delta \zeta$. To satisfy the variational problem we first require that $\int_{M_2} d^2z \, \sqrt{g^{(0)}} \, R_0$ fixed. This is nothing but the Euler character of the 2$d$ boundary $M_2$ and so is a topological condition. Using the traceless condition of $g^{(1)}_{ab}$ w.r.t $g^{(0)}_{ab}$ the left-over terms read:\footnote{This choice is a "non-chiral" way of solving the variational problem. There may be other "chiral" ways to do so that we do not consider.}
\bea
\delta g^{(1)}_{z \bar z} =0, ~~ g^{(1,1)}_{zz} = g^{(1,1)}_{\bar z \bar z} =0
\eea
We now turn to the implementation of the conditions $g^{(1,1)}_{zz} = g^{(1,1)}_{\bar z \bar z} =0$:
%

\bea
\label{condsonzeta}
-\kappa(z) = \frac{\partial_z^2 \zeta(z, \bar z)}{\zeta(z, \bar z)} - \frac{3 \, (\partial_z\zeta(z, \bar z))^2}{2 \, (\zeta(z, \bar z))^2}, ~~~ - \bar \kappa(\bar z) = \frac{\partial_{\bar z}^2 \zeta(z, \bar z)}{\zeta(z, \bar z)} - \frac{3 \, (\partial_{\bar z}\zeta(z, \bar z))^2}{2 \, (\zeta(z, \bar z))^2}.
\eea
However, for general $\zeta (z, \bar z)$ these equations cannot be imposed as the right hand side of $\kappa(z)$ ($\bar \kappa(\bar z)$) is not necessarily holomorphic (anti-holomorphic), and satisfying these conditions imposes further conditions on $\zeta (z, \bar z)$. To obtain what these conditions on $\zeta(z, \bar z)$ are, we start by noticing that the scalar curvature of the boundary metric $ds^2_{\Sigma_2} = \zeta(z, \bar z) \, dz \, d\bar z$ is
\bea
\label{r0forcg}
R_0 = \frac{4}{\zeta(z, \bar z)^3} \left[ \partial_z \zeta(z, \bar z) \, \partial_{\bar z} \zeta(z, \bar z) - \zeta(z, \bar z) \, \partial_z \partial_{\bar z} \zeta(z, \bar z) \right]
\eea
Let us note the following identities:
\bea
\partial_{\bar z} \left[\frac{\partial_z^2 \zeta(z, \bar z)}{\zeta(z, \bar z)} - \frac{3 \, (\partial_z\zeta(z, \bar z))^2}{2 \, (\zeta(z, \bar z))^2} \right] + \frac{1}{4} \zeta(z, \bar z) \, \partial_z R_0 &=& 0 \cr && \cr && \cr
\partial_z \left[\frac{\partial_{\bar z}^2 \zeta(z, \bar z)}{\zeta(z, \bar z)} - \frac{3 \, (\partial_{\bar z}\zeta(z, \bar z))^2}{2 \, (\zeta(z, \bar z))^2} \right] + \frac{1}{4} \zeta(z, \bar z) \, \partial_{\bar z} R_0 &=& 0 \, .
\eea
%
%
Thus to satisfy (\ref{condsonzeta}) we require that $R_0$ in (\ref{r0forcg}) is a constant. We need to consider the cases of $R_0 =0$ and $R_0 \ne 0$ separately.
\begin{itemize}
\item The case of $R_0=0$ means 
\bea
\partial_z \zeta(z, \bar z) \, \partial_{\bar z} \zeta(z, \bar z) = \zeta(z, \bar z) \, \partial_z \partial_{\bar z} \zeta(z, \bar z)
\eea
Writing $\zeta (z, \bar z) = e^{\phi(z, \bar z)}$ this equation is equivalent to $\partial_z \partial_{\bar z} \phi(z, \bar z) =0$ which is immediately solved by taking $\phi(z, \bar z) = \varphi(z) + \varphi(\bar z)$. Thus we write 
\bea
\zeta(z, \bar z) = f'(z) \bar f'(\bar z).
\eea
\item The case of $R_0 = r_0$ for non-zero constant $r_0$. Again writing $\zeta (z, \bar z) = e^{\phi(z, \bar z)}$ this reads:
\bea
\partial_z \partial_{\bar z} \phi(z, \bar z)+\frac{r_0}{4} e^{\phi(z, \bar z)} =0
\eea
\end{itemize}
This is the well-known Liouville equation on complex plane. And a general solution is %
\bea
\zeta (z, \bar z) = \frac{4 \alpha^2 f'(z) \, \bar f' (\bar z)}{(\beta + f(z) \, \bar f(\bar z))^2}
\eea
for which the curvature is $r_0 = 2 \beta/\alpha^2$. Whenever $r_0 >0$ we set $\alpha = \beta =1$ (which means we take $r_0=2$) and for $r_0<0$ we set $\alpha = -\beta =1$ (which corresponds to $r_0 =-2$). 

Let us also note that as a consequence of the consistent variational problem we have $\delta R_0 =0$.
%
This still leaves two terms proportional to $g_{(0)}^{ab} \partial_u g^{(2)}_{ab}$ and $D^a g^{(2)}_{au}$. Since we impose $g^{(1,1)}_{ab} =0$ we immediately have $\partial_u g^{(2)}_{ab}=0$ and we just have to impose 
\bea
D^a g^{(2)}_{au} = D^aD^b g^{(1,0)}_{ab} =0
\eea
This condition restricts the choice of $C(z, \bar z)$ in (\ref{lf-configs}). 
\begin{itemize}
\item For $\zeta = f'(z) \bar f'(\bar z)$.  
%
%
this equation can be solved in generality by writing:
\bea
\kappa(z) &=& \frac{3  f''(z)}{2 \, f'(z)^2} - \frac{ f'''(z)}{ f'(z)}, ~~~
\bar \kappa (\bar z) = \frac{3  \bar f''(\bar z)}{2 \, \bar f'(\bar z)^2} - \frac{ \bar f'''(\bar z)}{\bar f'( \bar z)} \cr &&  \cr
C(z, \bar z) &=& f_0(z) + \bar f_0(\bar z) + f(z) \, \bar f_1(\bar z) + \bar f (\bar z) \, f_1(z)
\eea
\item For $\zeta = (4 f'(z) \bar f'(\bar z))/(1+ f(z) \bar f(\bar z))^2$ we find that the solution to this constraint is:
\bea
\kappa(z) &=& \frac{3  f''(z)}{2 \, f'(z)^2} - \frac{ f'''(z)}{ f'(z)}, ~~~
\bar \kappa (\bar z) = \frac{3  \bar f''(\bar z)}{2 \, \bar f'(\bar z)^2} - \frac{ \bar f'''(\bar z)}{\bar f'( \bar z)} \cr &&  \cr
C(z, \bar z) &=& f_0(z) + \bar f_0(\bar z) + \frac{f(z) \, \bar f_1(\bar z) + \bar f (\bar z) \, f_1(z)}{1+ f(z) \bar f(\bar z)}
\eea
\end{itemize}
for arbitrary functions $(f_0(z), f_1(z), \bar f_0(\bar z), \bar f_1(\bar z))$. Thus in this case the locally flat solutions that satisfy variational problem is specified by three holomorphic functions $(f(z), f_0(z), f_1(z))$ and three anti-holomorphic functions $\bar f(\bar z), \bar f_0(\bar z), \bar f_1(\bar z))$. We do not pursue the symmetry algebra of this space of solution here.

\bibliographystyle{utphys}
\providecommand{\href}[2]{#2}\begingroup\raggedright\endgroup
\end{document}